\documentclass[fleqn,usenatbib]{mnras}

\usepackage{newtxtext,newtxmath}

\usepackage[T1]{fontenc}

\DeclareRobustCommand{\VAN}[3]{#2}
\let\VANthebibliography\thebibliography
\def\thebibliography{\DeclareRobustCommand{\VAN}[3]{##3}\VANthebibliography}

\usepackage{graphicx}	
\usepackage{amsmath}	
\usepackage{tikz}

\newcommand{\msun}{{\,\rm M_\odot}}

\newcommand{\kms}{\,{\rm km}\,{\rm s}^{-1}}

\newcommand{\erg}{\,{\rm erg}}

\newcommand{\Myr}{\,{\rm Myr}}

\newcommand{\Mpc}{\,{\rm Mpc}}

\def\jcap{J. Cosmol.  Astropart. Phys.}
\def\aap{A\&A}
\def\apj{ApJ}

\def\apjl{ApJ}
\def\mnras{MNRAS}
\def\araa{ARA\&A}
\def\aj{AJ}

\def\physrep{Phys. Rep.}
\def\nat{Nature}

\def\apjs{ApJS}
\def\prd{Phys. Rev. D}

\title[Early Galaxies and Early Dark Energy]{Early Galaxies and Early Dark Energy: A Unified Solution to the Hubble Tension and Puzzles of Massive Bright Galaxies revealed by JWST}

\author[Shen et al.]{\parbox{18.0cm}{
Xuejian Shen,$^{1}$\thanks{E-mail: xuejian@mit.edu}, 
Mark Vogelsberger,$^{1,2}$ 
Michael Boylan-Kolchin,$^{3}$
Sandro Tacchella,$^{4,5}$
Rohan P. Naidu$^{1}$\thanks{NASA Hubble Fellow}
}\vspace{0.3cm} \\
$^{1}$ Department of Physics \& Kavli Institute for Astrophysics and Space Research, Massachusetts Institute of Technology, Cambridge, MA 02139, USA \\
$^{2}$ The NSF AI Institute for Artificial Intelligence and Fundamental Interactions, Massachusetts Institute of Technology, Cambridge, MA 02139, USA \\
$^{3}$ Department of Astronomy, The University of Texas at Austin, 2515 Speedway Stop C1400, Austin, TX 78712, USA \\ 
$^{4}$ Kavli Institute for Cosmology, University of Cambridge, Madingley Road, Cambridge, CB3 0HA, UK \\
$^{5}$ Cavendish Laboratory, University of Cambridge, 19 JJ Thomson Avenue, Cambridge, CB3 0HE, UK \\
}

\date{Accepted XXX. Received YYY; in original form ZZZ}

\pubyear{2024}

\begin{document}
\label{firstpage}
\pagerange{\pageref{firstpage}--\pageref{lastpage}}
\maketitle

\begin{abstract}
JWST has revealed a large population of ultra-violet (UV)-bright galaxies at $z\gtrsim 10$ and possibly overly massive galaxies at $z\gtrsim 7$, challenging standard galaxy formation models in the $\Lambda$CDM cosmology. We use an empirical galaxy formation model to explore the potential of alleviating these tensions through an Early Dark Energy (EDE) model, originally proposed to solve the Hubble tension. Our benchmark model demonstrates excellent agreement with the UV luminosity functions (UVLFs) at $4\lesssim z \lesssim10$ in both $\Lambda$CDM and EDE cosmologies. In the EDE cosmology, the UVLF measurements at $z\simeq 12$ based on spectroscopically confirmed galaxies exhibit no tension with the benchmark model. Photometric constraints at $12 \lesssim z\lesssim 16$ can be fully explained within EDE via either moderately increased star formation efficiencies ($\epsilon_{\ast}\sim 3-10\%$ at $M_{\rm halo}\sim 10^{10.5}\msun$) or enhanced UV variabilities ($\sigma_{\rm UV}\sim 0.8-1.3$ mag at $M_{\rm halo}\sim 10^{10.5}\msun$) that are within the scatter of hydrodynamical simulation predictions. A similar agreement is difficult to achieve in $\Lambda$CDM, especially at $z\gtrsim 14$, where the required $\sigma_{\rm UV}$ exceeds the maximum value seen in simulations. Furthermore, the implausibly large cosmic stellar mass densities inferred from some JWST observations are no longer in tension with cosmology when the EDE is considered. Our findings highlight EDE as an intriguing unified solution to a fundamental problem in cosmology and the recent tensions raised by JWST observations. Data at the highest redshifts reached by JWST ($z \sim 14-16$) will be crucial for differentiating modified galaxy formation physics from new cosmological physics.
\end{abstract}




\section{Introduction}

The James Webb Space Telescope (JWST) has opened a new window for studying galaxy formation and evolution within the first $\sim 500\Myr$ ($z\gtrsim 10$) of the history of the Universe. Numerous photometric drop-out galaxy candidates at $z\gtrsim 9$ have been unveiled by the JWST/NIRCam imaging datasets~\citep[e.g.][]{Naidu2022,Castellano2022,Finkelstein2022,Adams2023,Atek2023,Bouwens2023a,Donnan2023,Harikane2023,Robertson2023,Yan2023,Hainline2024}, with unusually bright galaxy candidates revealed at $z\simeq 16$~\citep[e.g.][]{Harikane2023}. Among these galaxies, seven of the $z\gtrsim 12$ candidates have already been spectroscopically confirmed in the JADES~\citep{Robertson2023,Curtis2023,DEugenio2023,Carniani2024}, GLASS~\citep{Bakx2023,Castellano2024,Zavala2024}, and UNCOVER survey~\citep{Wang2023,Fujimoto2023b}.

The rest-frame UV luminosity functions (UVLFs) determined using JWST-identified galaxies show surprisingly little evolution at the bright end beyond $z\simeq 10$~\citep[e.g.][]{Harikane2023,Finkelstein2023} and differ substantially from the extrapolation of results derived from previous Hubble Space Telescope (HST) observations. Meanwhile, the number densities of UV-bright galaxies at $z\gtrsim 10$ inferred from these observations are in tension with predictions of the majority of the theoretical models~\citep[developed before JWST observations, as summarized in e.g.][]{Finkelstein2024,Finkelstein2023}. This includes empirical models~\citep[e.g.][]{Tacchella2013,Mason2015,Sun2016,Tacchella2018,Behroozi2020}, semi-analytical galaxy formation models~\citep[e.g.][]{Dayal2014,Dayal2019,Cowley2018,Yung2019,Mauerhofer2023,Yung2023}, and cosmological hydrodynamic simulations~\citep[e.g.][]{Simba,Vogelsberger2020,Haslbauer2022,Kannan2022-thesan,Kannan2022,Wilkins2023,Wilkins2023b}. This discrepancy is largely based on the photometrically-selected galaxy candidates, which could be contaminated by low-redshift interlopers~\citep[e.g.][]{Fujimoto2022, Zavala2023, Naidu2022b}. However, there has been a good agreement between the photometric and spectroscopic redshifts for the spectroscopically confirmed galaxies so far~\citep[e.g.][]{Finkelstein2024}. The pure spectroscopic constraints of the UVLF~\citep[e.g.][]{Harikane2024-spec,Harikane2024b-spec} also yield broadly consistent results with the photometric estimates. 

Such discrepancies are not completely unexpected, as many of these galaxy formation models were calibrated based on low-redshift observations, and qualitative differences in galaxy formation and evolution could exist in the extremely dense and low-metallicity environment of cosmic dawn~\citep[e.g.][]{Dekel2023,Ceverino2024,Lu2024}. Many physical interpretations of the tension have already been discussed in the literature, including but not limited to: (1) a substantially higher star-formation efficiency (SFE) in massive galaxies at $z\gtrsim 10$~\citep[e.g.][]{Li2023,Ceverino2024} potentially due to a feedback-free/failure regime~\citep[e.g.][]{Fall2010,Thompson2016,Grudic2018,Dekel2023,Menon2024}; (2) a top-heavy stellar initial mass function~\citep[IMF; e.g.][]{Inayoshi2022,Yung2023,Cueto2023,Trinca2024,Wang2024,Lu2024,Ventura2024} to increase the light-to-mass ratios of the stellar population, although increased feedback-to-mass ratios could cancel this effect~\citep[e.g.][]{Cueto2023}; (3) negligible dust attenuation~\citep[][but depends on the galaxy formation model used, as in some models, even the no-dust prediction is in tension with observations.]{Ferrara2022}; (4) UV radiation contributed by non-stellar sources, e.g. accreting stellar-mass black holes, quasars/active galactic nuclei~\citep[AGN; e.g.][]{Inayoshi2022,Trinca2024,Hegde2024} and see an observational case study of GN-z11~\citep{Tacchella2023}. 

The solutions above focus on enhancing the UV photon yield in high-redshift galaxies. An alternative and orthogonal solution involves variability/stochasticity of galaxy UV luminosity at fixed halo mass~\citep[e.g.][]{Mason2023,Mirocha2023,Shen2023,Sun2023,Kravtsov2024,Gelli2024} without changing the median UV photon yield of galaxies. In this scenario, the bright end of the UVLF is populated by a large number of low-mass haloes with upscattered UV brightness. In \citet{Shen2023}, a constant UV variability of $\sigma_{\rm UV}\simeq 1.5\,(2.5)$ mag is suggested to reconcile observations at $z\simeq 12\,(16)$. The $\sigma_{\rm UV}$ value is reduced to $\simeq 1 - 1.3$ (2) mag with more detailed modelling of the bursty star-formation histories of galaxies~\citep{Kravtsov2024}. \citet{Gelli2024} considered the halo mass-dependence of $\sigma_{\rm UV}$ and found that the $z\simeq 12$ observations can be reconciled with theoretical models without modifications to $\sigma_{\rm UV}(M_{\rm halo})$ at lower redshifts, but the $z\gtrsim 14$ results remain challenging.

Observational signatures of large burstiness of star-formation in high-redshift galaxies have been found~\citep[e.g.][]{Ciesla2023,Cole2023,Endsley2023,Looser2023,Tacchella2023b,Dressler2024,Helton2024} with a potential qualitative transition around $z\simeq 10$. The major source of this variability could come from the bursty star-formation in low-mass, high-redshift galaxies, often seen in cosmological zoom-in simulations with advanced models for star-formation and feedback in the interstellar medium~\citep[ISM; e.g.][]{Hopkins2018, FG2018, Marinacci2019,Sun2023, Katz2023}. Increased variability at $z\gtrsim 10$ could come from stochastic gas inflow~\citep[e.g.][]{Tacchella2020}, rapid star-formation in a feedback-failure regime~\citep[e.g.][]{Grudic2018,Dekel2023,Menon2024}, in massive star clusters due to Lyman-Werner radiation feedback in metal-poor environment~\citep[e.g.][]{Sugimura2024}, or due to highly clustered (in both space and time) feedback from a top-heavy IMF. Enhanced variability in clumpy, high-redshift galaxies could also result indirectly from the limited statistics of star-forming regions and the increased sampling noises.

However, the aforementioned ideas all assume the standard $\Lambda$CDM cosmology. While this model has been singularly successful at explaining a variety of cosmological observations across a range of length scales and cosmic epochs, questions still remain about whether it is a complete (albeit phenomenological) description of the evolution of the Universe. In particular, a persistent discrepancy between inferences of the current expansion rate of the Universe based on modeling anisotropies in the Cosmic Microwave Background (CMB; e.g. \citealt{Planck2020}) and directly measuring the expansion locally~\citep[e.g.][]{Riess2022} --- the so-called ``Hubble tension'' (see \citealt{Abdalla2022} for a review) --- has prompted a flurry of activity aimed at understanding how to preserve $\Lambda$CDM's successes while resolving this tension (see \citealt{DiValentino2021} for a review). 

Given the abundance of precision cosmological data now available, modifications to the standard $\Lambda$CDM that do not conflict with observations are surprisingly difficult to realize. The ``least unlikely'' type of modification that is allowed~\citep{Knox2020} increases the expansion rate of the Universe before recombination. This modification decreases the physical sound horizon measured by the CMB, requiring an attendant decrease in the distance to the last scattering that is obtained through an increase in $H_0$. One possible mechanism for achieving this increased expansion at early times is ``Early Dark Energy'' (EDE), where a new cosmological energy source with an equation of state similar to dark energy contributes $\sim 10\%$ of the critical density at its time of maximal contribution ($z \sim 3000$; see \citealt{Poulin2023} for a recent review). In this work, we specifically explore a scalar field model of EDE, as proposed in \citet{Karwal2016, Poulin2018, Poulin2019, Smith2020} to solve the Hubble tension. As the EDE in this model decays rapidly after recombination and exerts negligible dynamical effects at late times, its primary influence on structure formation is through altered cosmological parameters, notably increasing both the amplitude $A_{\rm s}$ and spectral index $n_{\rm s}$ of primordial scalar fluctuations, and the physical matter density $\omega_{\rm cdm}$.

As a consequence of these changes in cosmological parameters, the halo and thus galaxy abundance are systematically enhanced at high redshifts~\citep[e.g.][]{Klypin2021,BK2023,Forconi2024}, which has the potential to alleviate the UVLF tension discussed above. The enhanced halo abundance also has implications for another tension raised by JWST regarding potentially overly-massive galaxies at $z\gtrsim 5$~\citep[e.g.][]{Labbe2023,Akins2023,Xiao2023,deGraaff2024,Casey2024}. In some cases~\citep[e.g.][]{Labbe2023}, the implied cosmic stellar mass density exceeds the total mass budget of baryons in a $\Lambda$CDM universe~\citep{BK2023,Lovell2023}. Despite the debated nature of these sources~\citep[e.g.][]{Endsley2022,Larson2022,Kocevski2023,Desprez2024,Narayanan2024,Wang2024}, EDE offers an alternative way to alleviate this potential cosmological tension.

In this paper, we investigate the potential of EDE as a unified solution to the Hubble tension and the tensions regarding UV-bright and potentially overly massive galaxies raised by JWST. The paper is organized as follows: In Section~\ref{sec:cosmology}, we introduce the EDE model and the calculation of dark matter halo mass functions as well as growth rates. In Section~\ref{sec:galform}, we discuss the galaxy formation model and establish a median mapping between the halo mass function and galaxy UVLF. We then describe how we treat UV variability and its halo mass dependence. In Section~\ref{sec:results}, we present the results and discuss how EDE changes the landscape of UVLFs at $z\gtrsim 10$ and the stellar mass densities at $z\gtrsim 7$. In Section~\ref{sec:conclusions}, we provide discussions and our conclusions.

\begin{table}
\centering
\addtolength{\tabcolsep}{8pt}
\def\arraystretch{1.2}
\begin{tabular}{lll}
\hline
\hline
Model & $\Lambda$CDM & EDE \\
\hline
$f_{\rm EDE}(z_{\rm c})$ & - & $0.179^{+0.047}_{-0.04}$ \\
$\log_{10}(z_{\rm c})$ & - & $3.528^{+0.028}_{-0.024}$ \\
$\theta_{\rm i}$ & - & $2.806^{+0.098}_{-0.093}$ \\
$m$ ($10^{-28}$eV) & - & $4.38 \pm 0.49$ \\
$f$ (Mpl) & - & $0.213 \pm 0.035$ \\
\hline
$H_0$ [$\kms\Mpc^{-1}$] & $67.81^{+0.64}_{-0.6}$ & $74.83^{+1.9}_{-2.1}$ \\
$100\omega_b$ & $2.249^{+0.014}_{-0.013}$ & $2.278^{+0.018}_{-0.02}$ \\
$\omega_{\rm cdm}$ & $0.1191^{+0.0014}_{-0.0015}$ & $0.1372^{+0.0053}_{-0.0059}$ \\
$10^9A_{\rm s}$ & $2.092^{+0.035}_{-0.033}$ & $2.146^{+0.041}_{-0.04}$ \\
$n_{\rm s}$ & $0.9747^{+0.0046}_{-0.0047}$ & $1.003^{+0.0091}_{-0.0096}$ \\
$S_8$ & $0.821 \pm 0.017$ & $0.829^{+0.017}_{-0.019}$ \\
$\Omega_{\rm m}$ & $0.309^{+0.009}_{-0.008}$ & $0.287 \pm 0.009$ \\
\hline
\end{tabular}
\caption{The best-fit $\pm 1\sigma$ errors of the cosmological parameters reconstructed in the $\Lambda$CDM and EDE models from the analysis of the ACT DR4 + SPT-3G+Planck TT650TEEE dataset combination in \citet{Smith2022}. $f_{\rm EDE}(z) \equiv \Omega_{\rm a}(z)/\Omega_{\rm tot}(z)$ is the fraction of energy density contributed by EDE. $z_{\rm c}$ is the critical redshift when EDE becomes dynamical. $\theta_{\rm i}$ is the initial field value before oscillation. $m$ is the mass of the scalar field. $f$ is the decaying constant of the field in unit of Planck scale (Mpl). $H_{\rm 0}$ is the Hubble constant at $z=0$. $\omega_{\rm x} \equiv \Omega_{x}\,h^{2}$, where $h\equiv H_{0}/100$. $A_{\rm s}$ and $n_{\rm s}$ are the normalization and power-law index of the primordial power spectrum. $\Omega_{\rm m}$ is matter density. $S_{8}\equiv \sigma_{8} (\Omega_{\rm m}/0.3)^{1/2}$.}
\label{tab:parameters}
\end{table}

\begin{figure}
    \centering
    \includegraphics[width=0.96\linewidth]{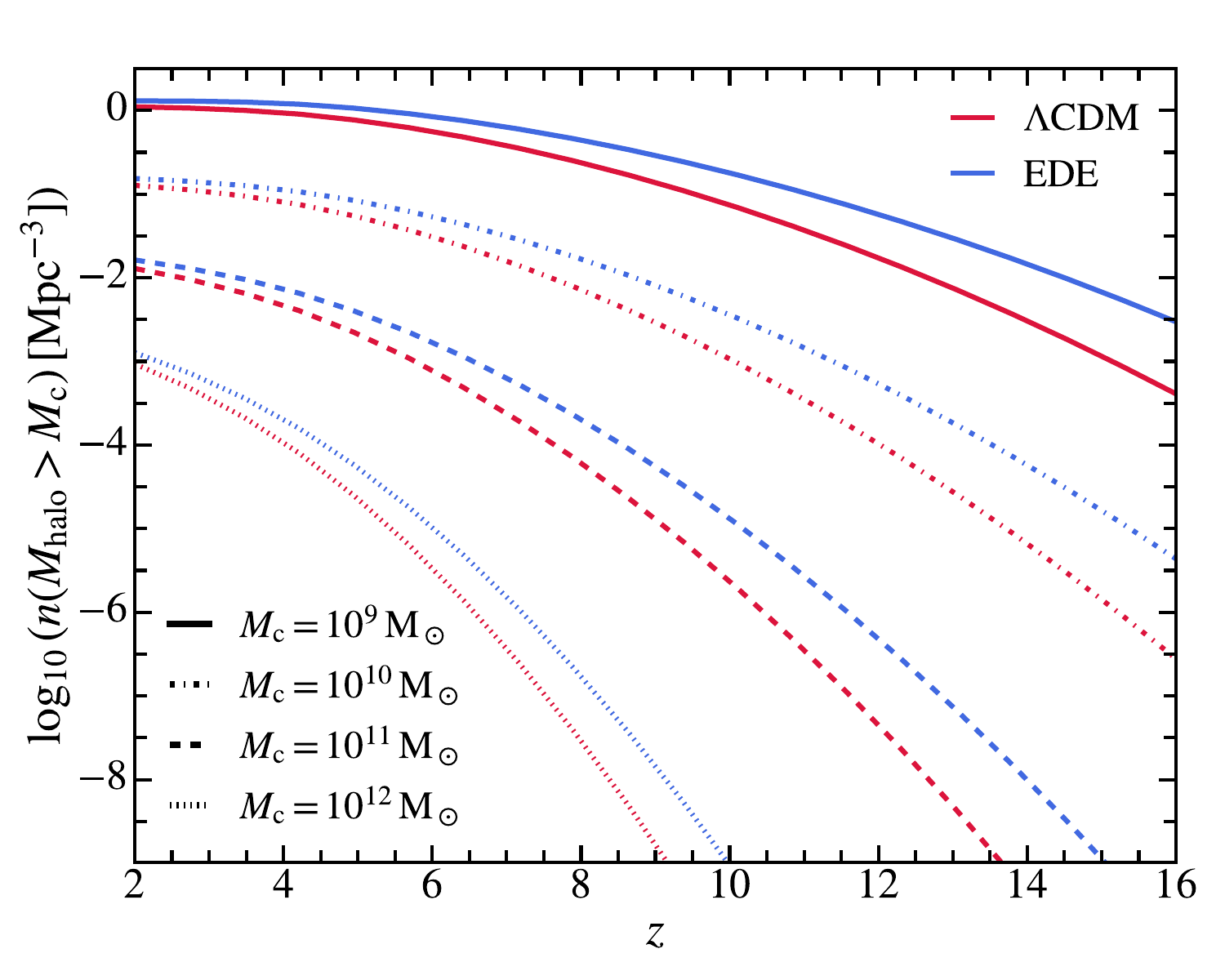}
    \caption{Number density of haloes above mass $M_{\rm c}$ versus redshift in the EDE and $\Lambda$CDM cosmology. We show results with $M_{\rm c}=10^{9}, 10^{10}, 10^{11}, 10^{12}$ with different types of lines. The halo counts in the EDE model are systematically enhanced at all mass scales at $z\gtrsim 8$. However, the differences diminish at low redshifts.}
    \label{fig:nhalo}
\end{figure}

\section{Cosmological model}
\label{sec:cosmology}

The EDE considered in this paper is a scalar field with an axion-like potential $V_{\rm n}(\phi)\sim [1-{\rm cos}(\phi/f)]^{\rm n}$, where $f$ is the decay constant of the field. These ultralight axion-like fields arise generically in string theory~\citep[e.g.][]{Svrcek2006,Arvanitaki2010,Kamionkowski2014,Marsh2016} and have intriguing cosmological implications for dark matter~\citep{Marsh2016} and EDE~\citep{Poulin2019}. The field is frozen at early times and acts as a cosmological constant. The field becomes dynamical at a critical redshift $z_{\rm c}$ as the Hubble friction decreases, eventually settling down around the minimum of the potential and starting oscillation. The effective equation-of-state of the field afterward is $w_{\rm n} \simeq (n-1)/(n+1)$. Here following \citet{Smith2022}, we consider the $n=3$ case, where the energy density of EDE dilutes faster than radiation. The existence of this additional pre-recombination energy density increases the Hubble parameter around the time of photon decoupling, and reduces the physical sound horizon if $H_0$ is fixed. Therefore, the inferred $H_0$ from the same sound horizon measurement on CMB will be increased with EDE. Since the energy density of the EDE field dilutes rapidly, it casts no direct impact on structure formation but its effects are indirectly imprinted through changes in cosmological parameters. We choose EDE and cosmological parameters as the best-fit values in \citet{Smith2022} constrained jointly by ACT, SPT, and Planck results. They are summarized in Table~\ref{tab:parameters} along with the best-fit parameters in the standard $\Lambda$CDM cosmology. The best-fit $H_0$ in the EDE cosmology is around $74\kms \Mpc^{-1}$,  in agreement with the local constraints.

\begin{figure}
    \centering
    \includegraphics[width=\linewidth]{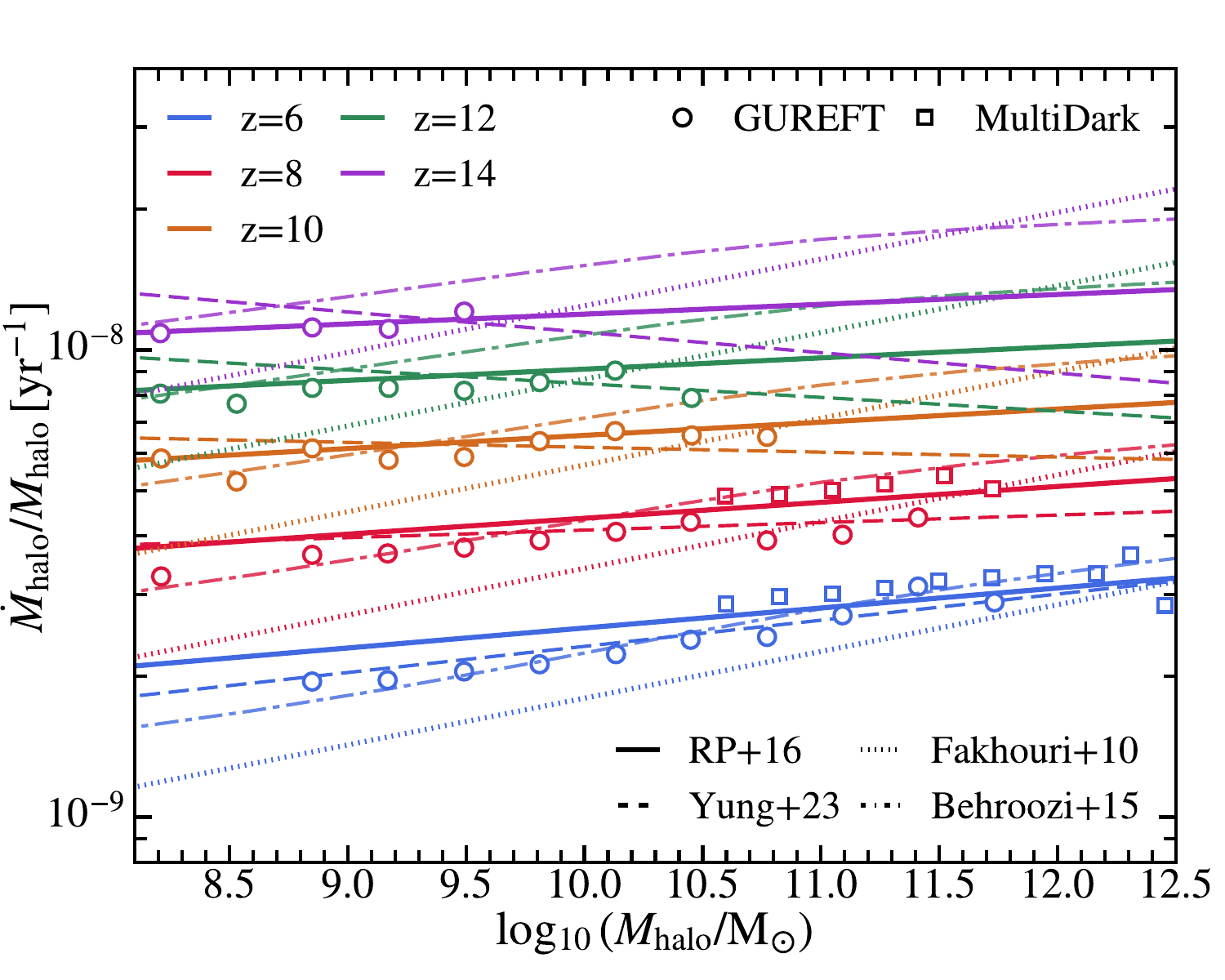}
    \caption{Halo specific accretion rate versus halo mass. We compare the halo accretion rates in \citet{Fakhouri2010} (adopted in \citealt{Shen2023}) with the fitting functions in \citet{Behroozi2015}, \citet{RP2016} (RP16), and \citet{Yung2024}. We also show the simulation results in \textsc{Gureft}~\citep{Yung2024} and MultiDark simulations~\citep{Klypin2016}. \citet{Behroozi2015} gives the mean accretion rate of $M_{\rm peak}$ and we correct it to the median value by assuming a $0.3$ dex scatter in halo accretion rates~\citep[e.g.][]{Rodrguez2016, Ren2019, Mirocha2021}. It tends to overestimate the halo accretion rates in the massive end at $z\gtrsim 10$. The \citet{Fakhouri2010} model underestimates the accretion rates in the low-mass end. The fitting in \citet{Yung2024} is restricted to relatively low-mass haloes at high redshifts and it is unclear whether the fitted negative slope holds at $z\gtrsim 10$. Therefore, we choose \citet{RP2016} as our new fiducial model.}
    \label{fig:halo_accretion}
\end{figure}

\subsection{Halo mass function}

The halo mass function is constructed following Press-Schechter-like theories \citep[e.g.][]{Press1974, Bond1991, Sheth2001} as implemented in the {\sc Hmf} code~\citep{hmf2,hmf3}. The transfer function is calculated using the Code for Anisotropies in the Microwave Background (\textsc{Camb}; \citealt{CAMB1,CAMB2}) and specifically the axion effective fluid model\footnote{We have also experimented with the early quintessence model~\citep{Smith2020} implemented in \textsc{Camb} and find $\lesssim 0.01$ dex differences between the halo mass function at $z=10$ calculated using the transfer functions from the two EDE model implementations.}~\citep{Poulin2018} implemented there. Following \citet{Planck2020}, in both $\Lambda$CDM and EDE, we model free-streaming neutrinos as two massless and one massive species with $m_{\nu} = 0.06\,{\rm eV}$. The effective number of neutrino species $N_{\rm eff}$ is set to $3.046$. We adopt a real-space top-hat filter function for the density field. The definition of halo mass follows the virial criterion in \citet{Bryan1998}. We adopt the halo mass function parametrization of \citet{Behroozi2013} to improve the accuracy at high redshift. 

In Figure~\ref{fig:nhalo}, we show the cumulative number density of haloes above a certain mass threshold as a function of redshift in the EDE and $\Lambda$CDM cosmology. The number densities of haloes are enhanced in the EDE model preferentially at high redshifts and at all mass scales because of the higher values of $n_{\rm s}$, $\omega_{\rm cdm}$ and $\sigma_8$ in EDE relative to the Planck $\Lambda$CDM cosmology~\citep[e.g.][]{Klypin2021,BK2023,Forconi2024}. By $z=0$, the halo mass functions are indistinguishable.

\subsection{Halo accretion rate}

We use the fitting function of halo accretion rate in \citet{RP2016}
\begin{align}
    &\dot{M}_{\rm halo} = C \, \left( \dfrac{M_{\rm halo}}{10^{12}\msun/h} \right)^{\gamma} \, \dfrac{H(z)}{H_{0}}, \\
    &\gamma = 1.000 + 0.329\, a - 0.206\,a^2, \nonumber \\
    &\log_{10}{C} = 2.730 - 1.828 \, a + 0.654\, a^2, \nonumber 
\end{align}
where $a = 1/(1+z)$ is the scale factor. The $\dot{M}_{\rm halo}$ here is the median value at a given halo mass and is averaged over one dynamical time of the halo. This relation is calibrated on the Bolshoi-Planck and MultiDark-Planck cosmological simulations~\citep{Klypin2016}. In Figure~\ref{fig:halo_accretion}, we compare this fitting formula with relations found in other works~\citep{Fakhouri2010,Behroozi2015,Yung2023} and N-body simulation results~\citep{Klypin2016,Yung2023}. We find that the fitting function from \citet{RP2016} gives better agreement to simulations at high redshifts over a large dynamical range and therefore choose it as our fiducial model. We also note that the dependence of $\dot{M}_{\rm halo}$ on cosmology is fully absorbed in the $H(z)$ term and we expect it to hold in the EDE cosmology considered here as well.

\subsection{Additional cosmological corrections}
For predictions in the EDE model, the ``interpreted'' galaxy luminosities and number densities by an observer assuming $\Lambda{\rm CDM}$ should have additional corrections as
\begin{align}
    & \Phi^{\prime} = \Phi  \times  \dfrac{({\rm d}V/{\rm d}z)_{\rm EDE}}{({\rm d}V/{\rm d}z)_{\Lambda{\rm CDM}}}, \nonumber \\
    & M_{\rm UV}^{\prime} = M_{\rm UV} + 2.5 \log_{10}{\left( (D^{\rm EDE}_{\rm L} / D^{\Lambda{\rm CDM}}_{\rm L})^{2}  \right)},
\end{align}
where $({\rm d}V/{\rm d}z)$ and $D_{\rm L}$ are the differential comoving volume and luminosity distance at the redshift of interest. The galaxy stellar masses, luminosities, and number densities in the EDE cosmology will all be corrected values throughout this paper to form better comparisons with observational data.

\section{Empirical galaxy formation model}
\label{sec:galform}

\begin{figure}
    \centering
    \includegraphics[width=\linewidth]{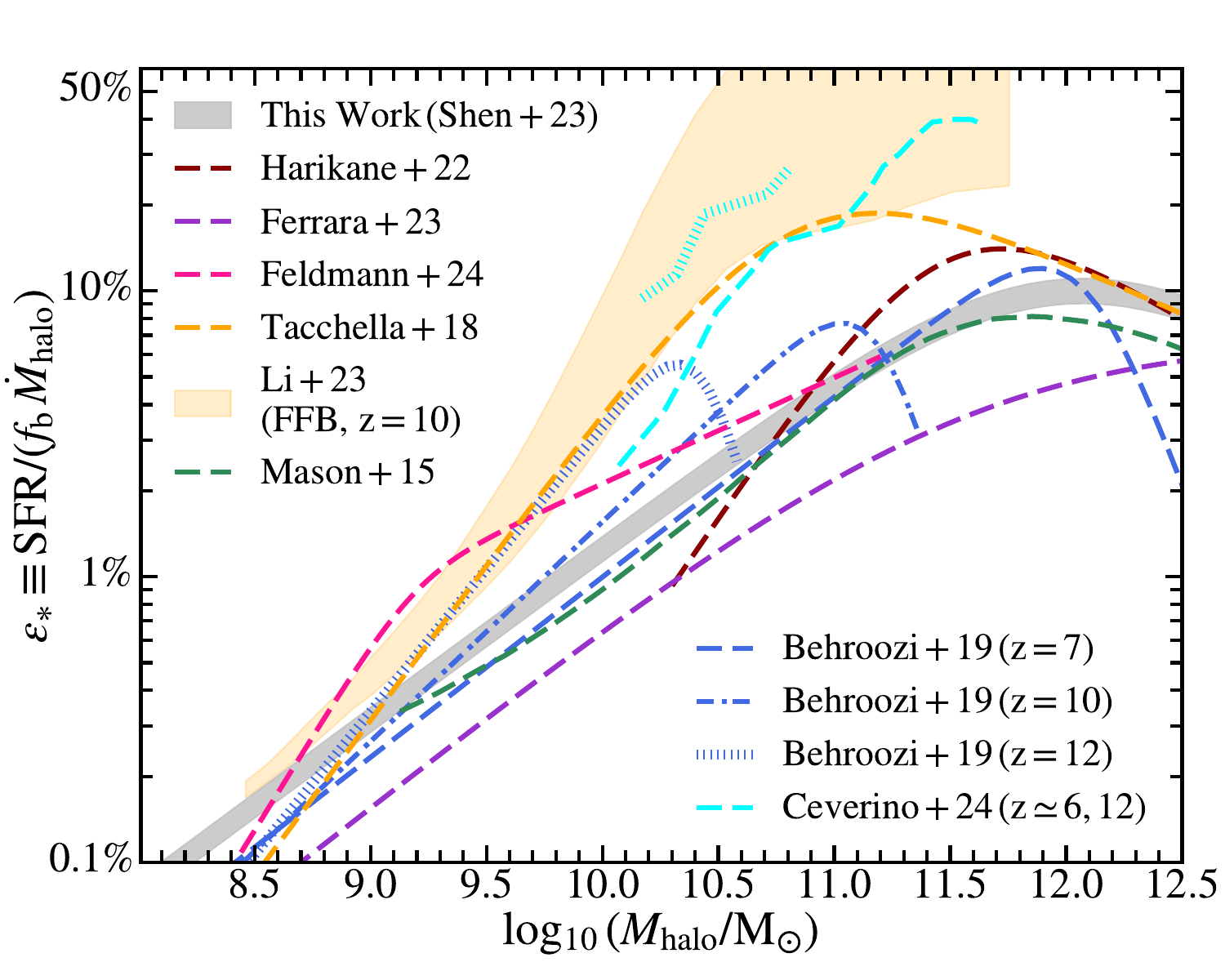}
    \caption{Star-formation efficiency (SFE) versus halo mass. We show the relation adopted in this work with the gray band, varying $\epsilon$ by $\pm 10\%$. We compare it with the SFE in other works~\citep[][]{Mason2015,Tacchella2018,Behroozi2019,Harikane2022,Ferrara2023,Li2023,Ceverino2024,Feldmann2024} as labelled. For the \textsc{Universe Machine}~\citep{Behroozi2019} and \citet{Ferrara2023} predictions, we take their SFR-$M_{\rm halo}$ relations and convert it to SFE assuming our halo accretion rate model. For the FIREbox~\citep{Feldmann2024} predictions, we show the SFE based on the averaged SFR in the 100-Myr window. \citet{Li2023} adopted a modified version of \citet{Behroozi2019} with feedback-free starburst (FFB) in massive haloes. Results from the FirstLight simulations~\citep{Ceverino2024} show signatures of this regime at $z\simeq 12$ (dotted cyan line). For the observational determination in \citet{Harikane2022}, we scale it down to account for the different IMF choices.}
    \label{fig:sfe}
\end{figure}

\begin{figure}
    \centering
    \includegraphics[width=\linewidth]{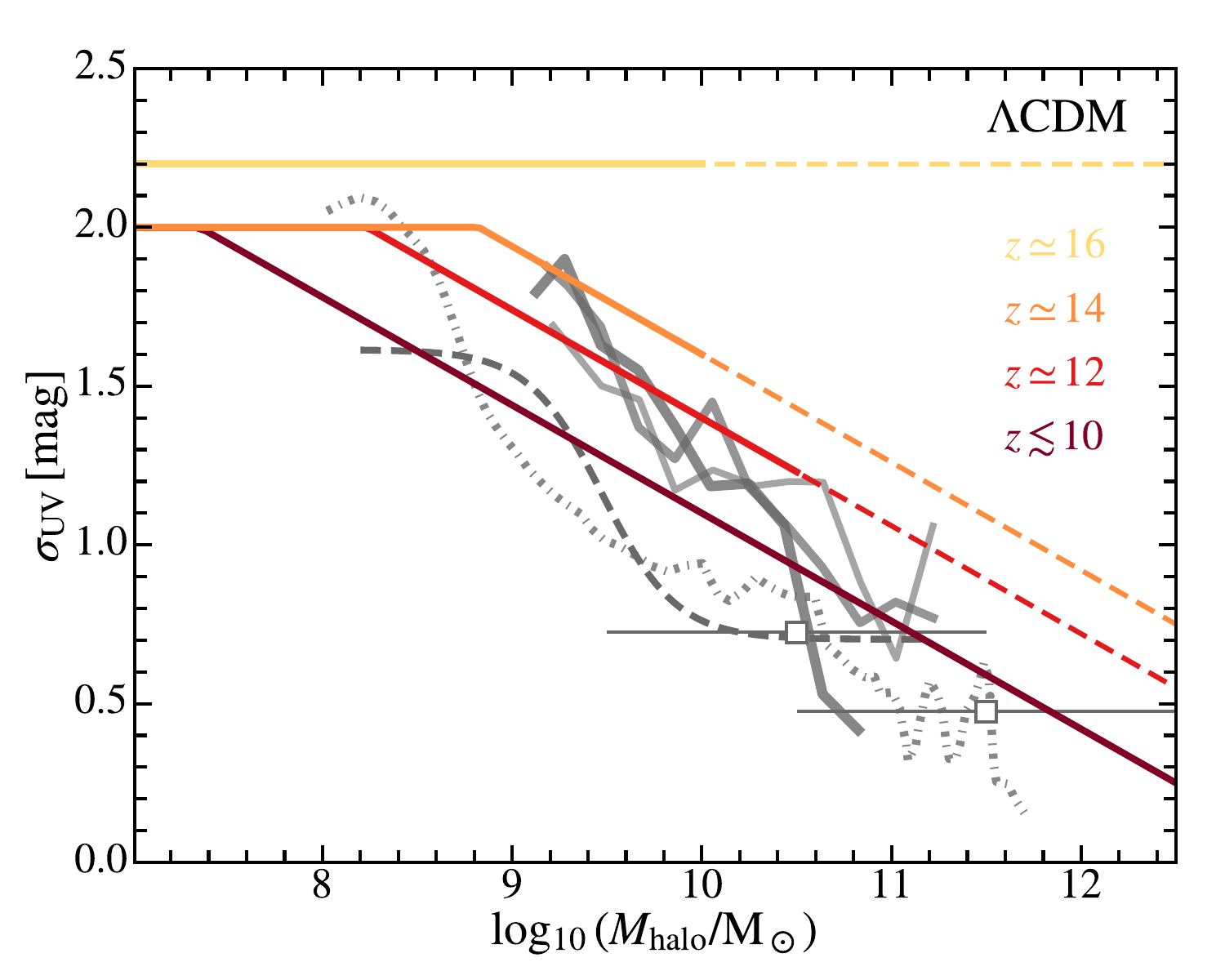}
    \includegraphics[width=\linewidth]{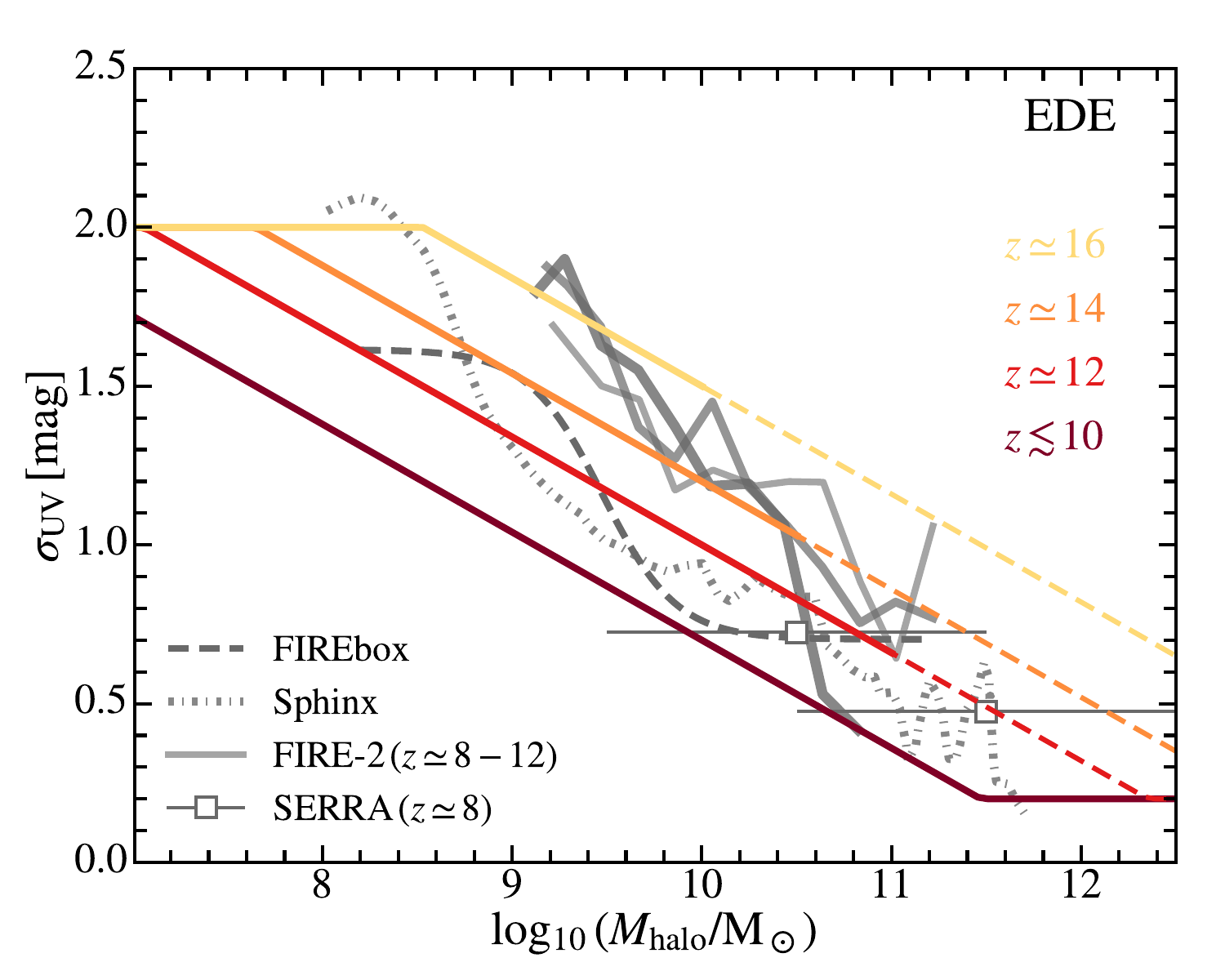}
    \caption{UV variability $\sigma_{\rm UV}$ versus $M_{\rm halo}$. We show the parametric model adopted in this work, which roughly scales as $M_{\rm halo}^{-1/3}$. Different values of the normalization factor $A$ are preferred at different redshifts beyond $\sim 10$. We compare it to the results from FIRE-2 zoom-in simulations~\citep{Sun2023}, the FIREbox simulation~\citep{Feldmann2024}, the \textsc{Sphinx} simulations~\citep{Katz2023} as shown in \citet{Kravtsov2024}, and the \textsc{Serra} simulations~\citep{Pallottini2023}. In both $\Lambda$CDM and EDE, a fixed $\sigma_{\rm UV}$-$M_{\rm halo}$ relation is capable of matching observational results at $z\lesssim 10$. However, at $z\gtrsim 12$, an enhancement of $\sigma_{\rm UV}$ is required and becomes prohibitively high at $z\simeq 16$. In EDE, similar relative enhancement in $\sigma_{\rm UV}$ is inferred at $z\simeq 12 - 14$ but the overall values of $\sigma_{\rm UV}$ agree reasonably with theoretical predictions. We note that the UVLFs at $z\gtrsim 10$ in the luminosity range probed by observations will not be sensitive to the variability in the massive end, which is indicated by the dashed lines.}
    \label{fig:sigma_vs_mhalo}
\end{figure}

\begin{figure*}
    \centering
    \begin{tikzpicture}

        \node at (0.6\textwidth,0.8\textwidth) {\includegraphics[width=0.66\textwidth]{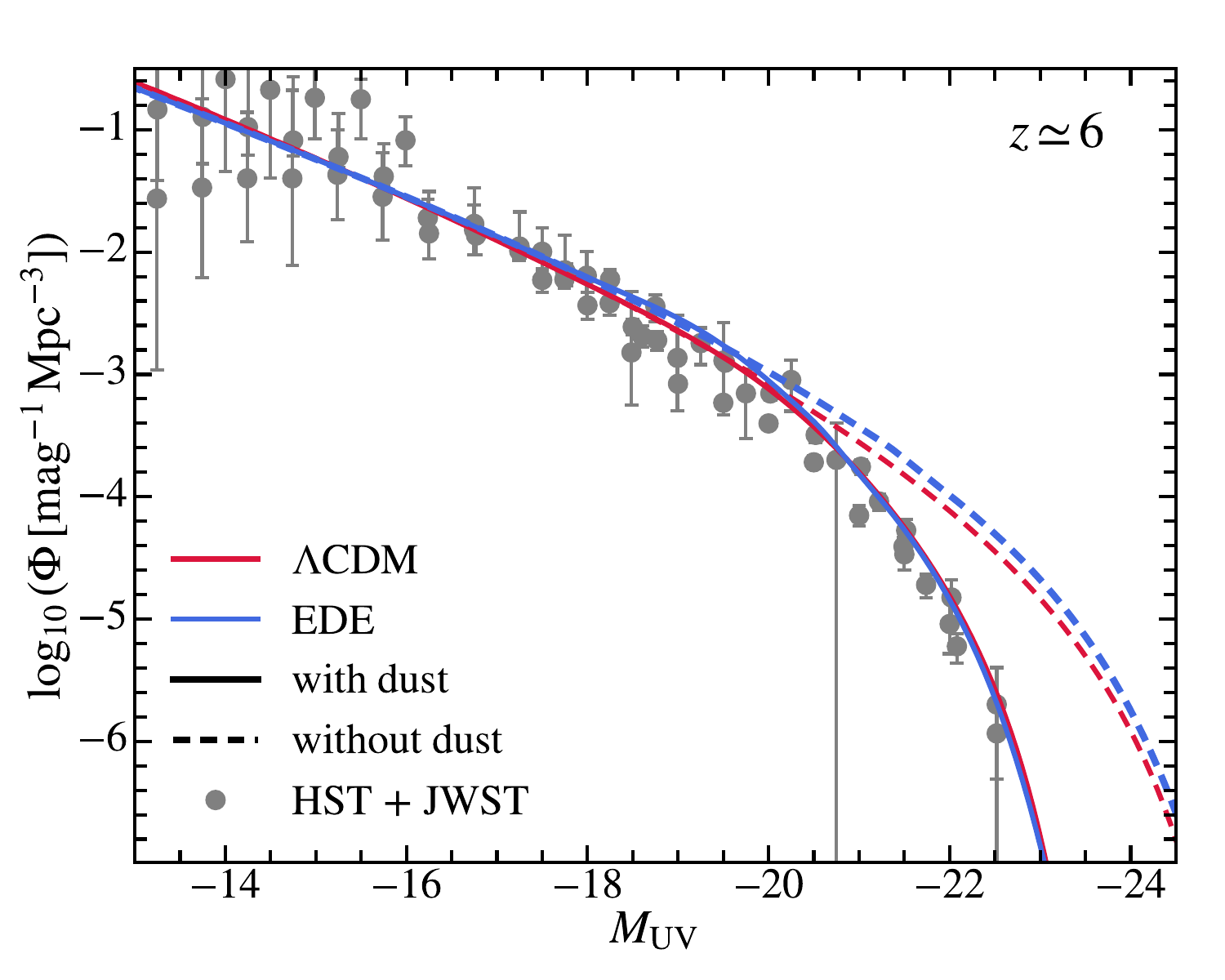}};
        \node at (0.1\textwidth,0.91\textwidth) {\includegraphics[width=0.33\textwidth]{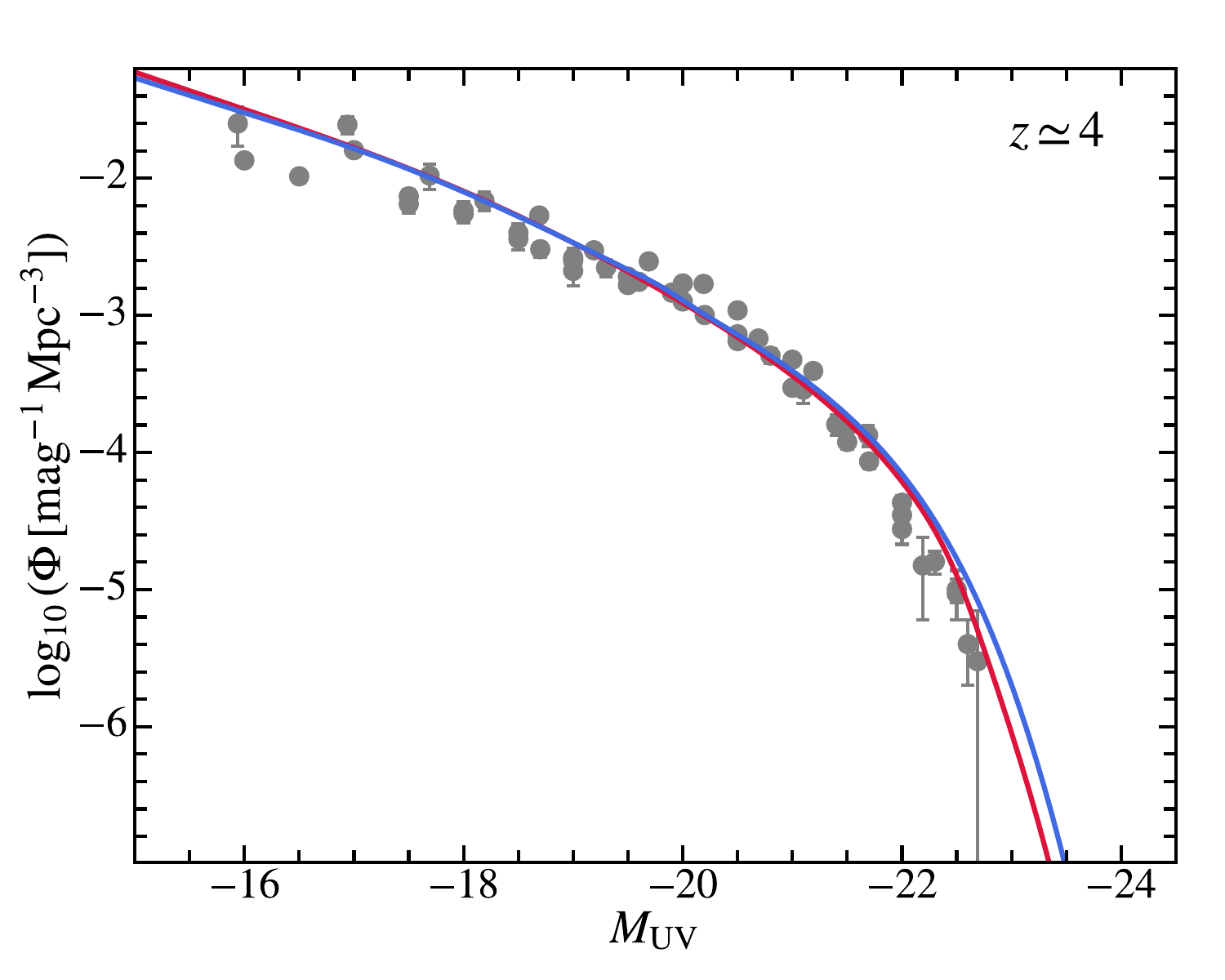}};
        \node at (0.1\textwidth,0.665\textwidth) {\includegraphics[width=0.33\textwidth]{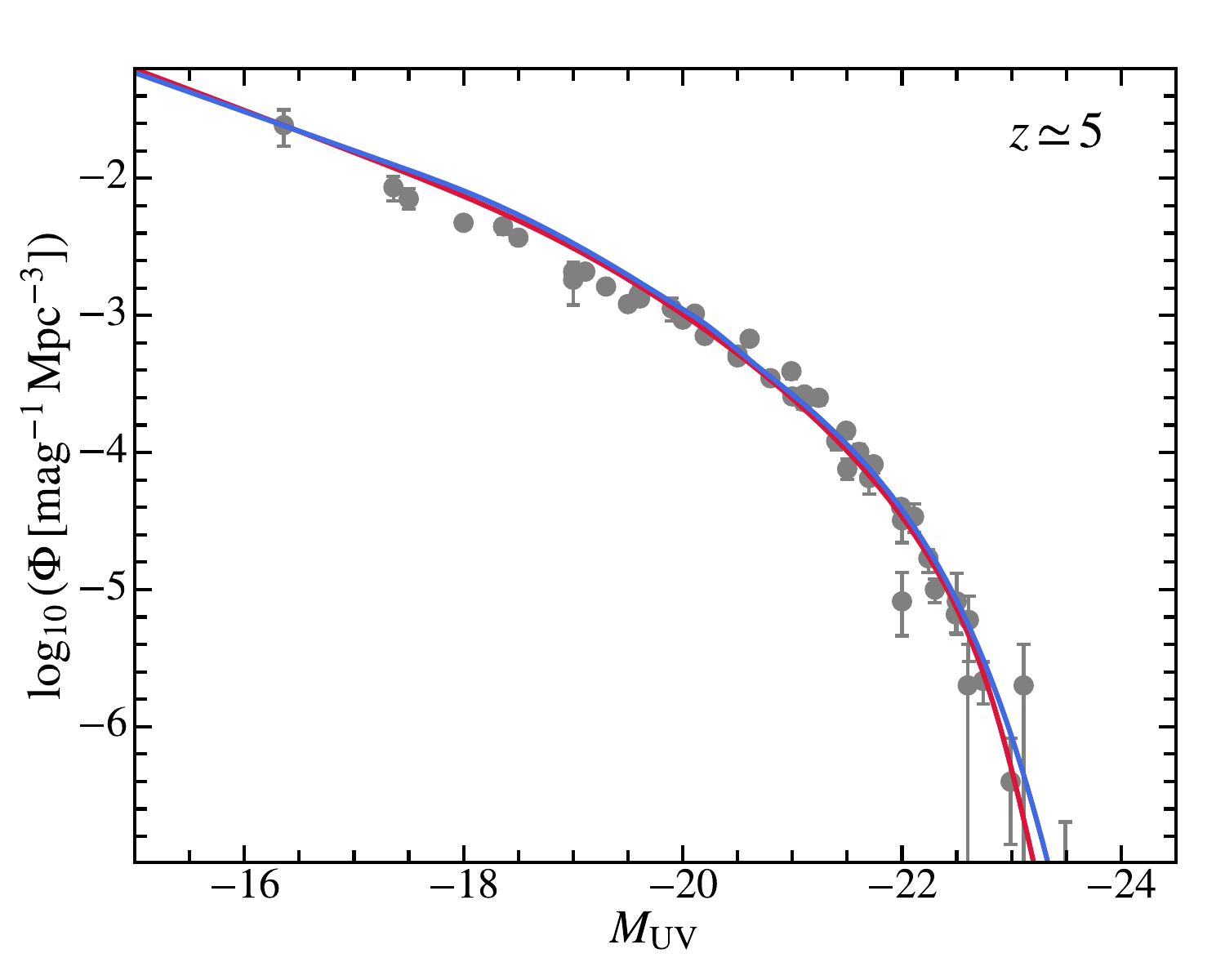}};
        \node at (0.1\textwidth,0.42\textwidth) {\includegraphics[width=0.33\textwidth]{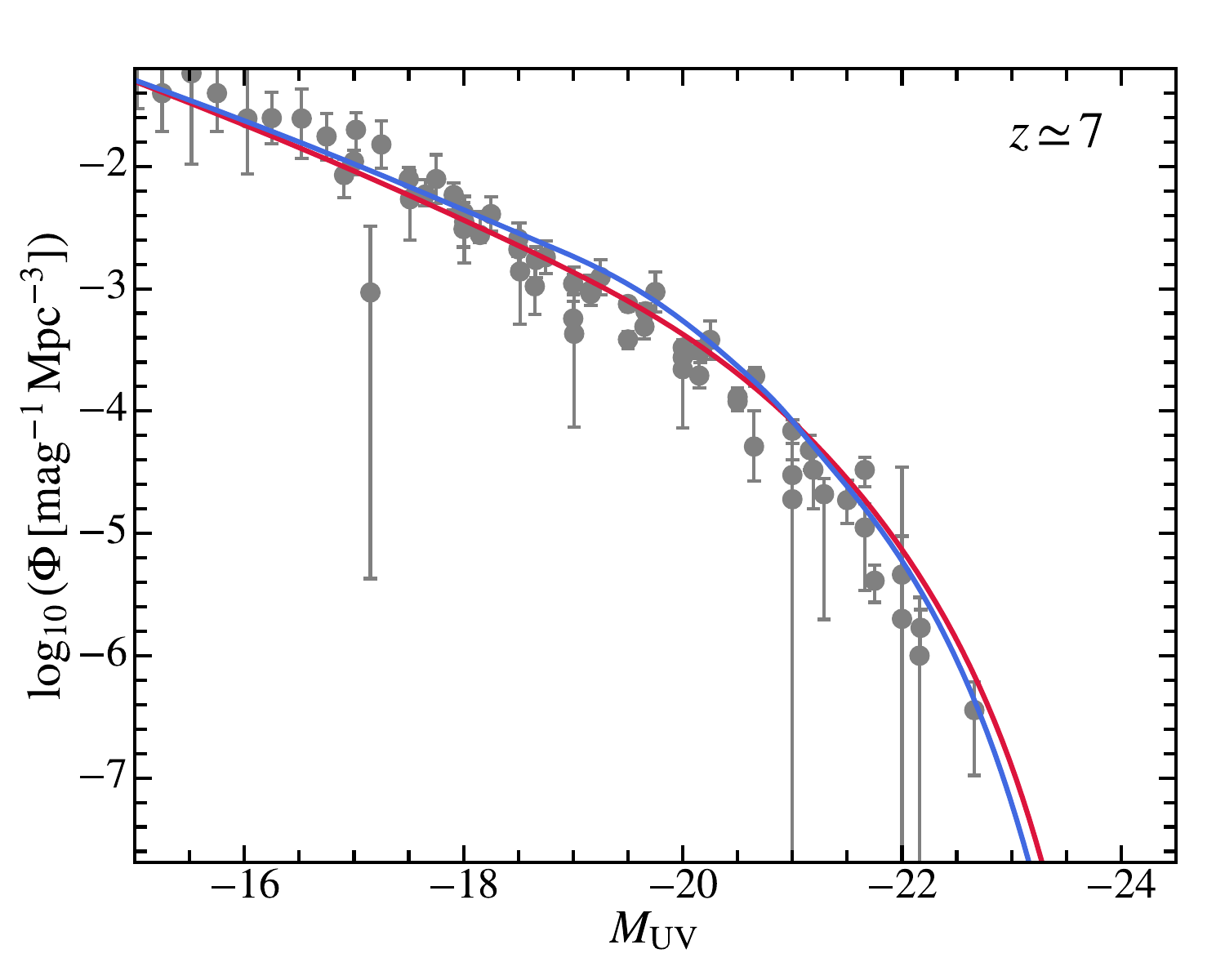}};
        \node at (0.43\textwidth,0.42\textwidth) {\includegraphics[width=0.33\textwidth]{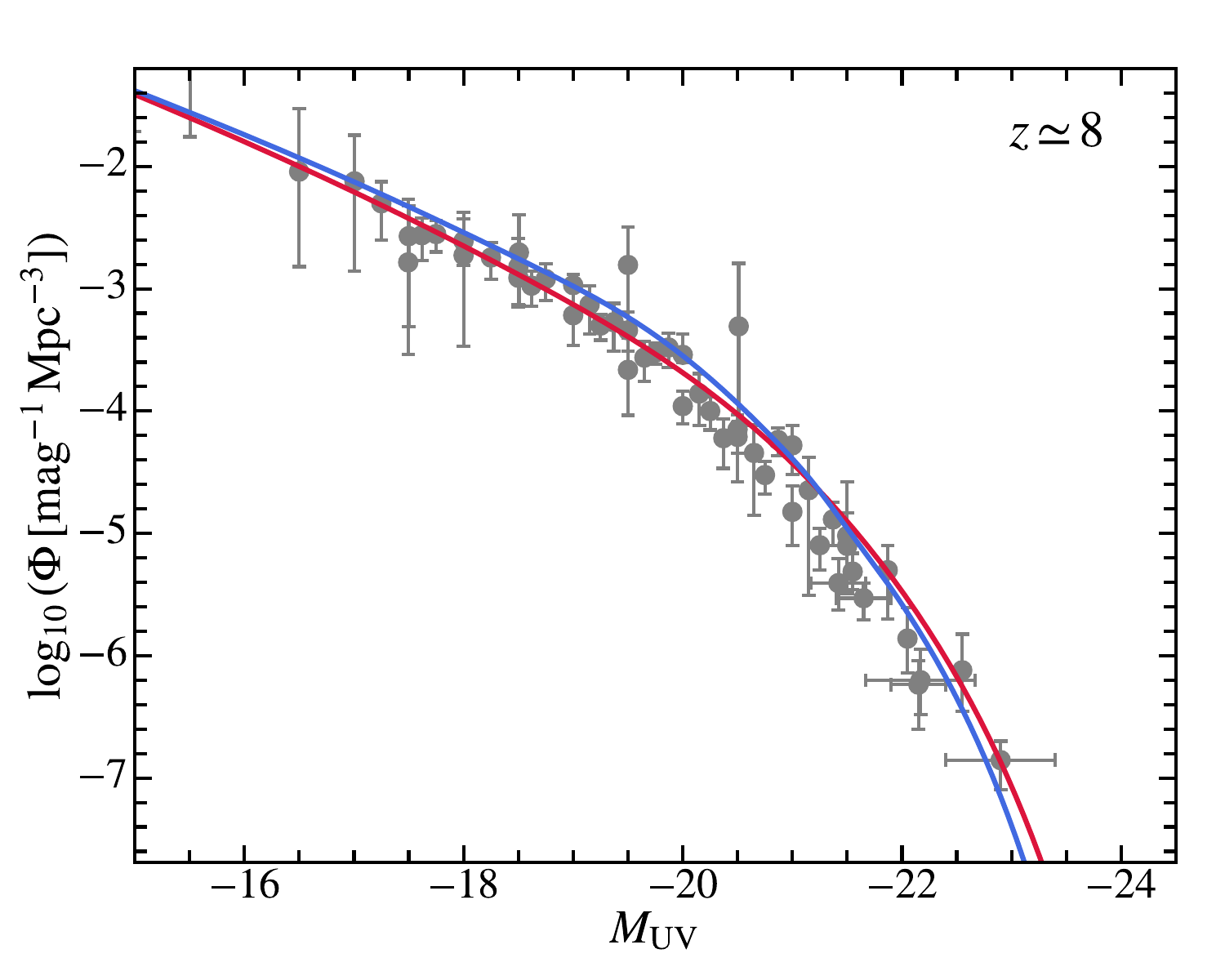}};
        \node at (0.76\textwidth,0.42\textwidth) {\includegraphics[width=0.33\textwidth]{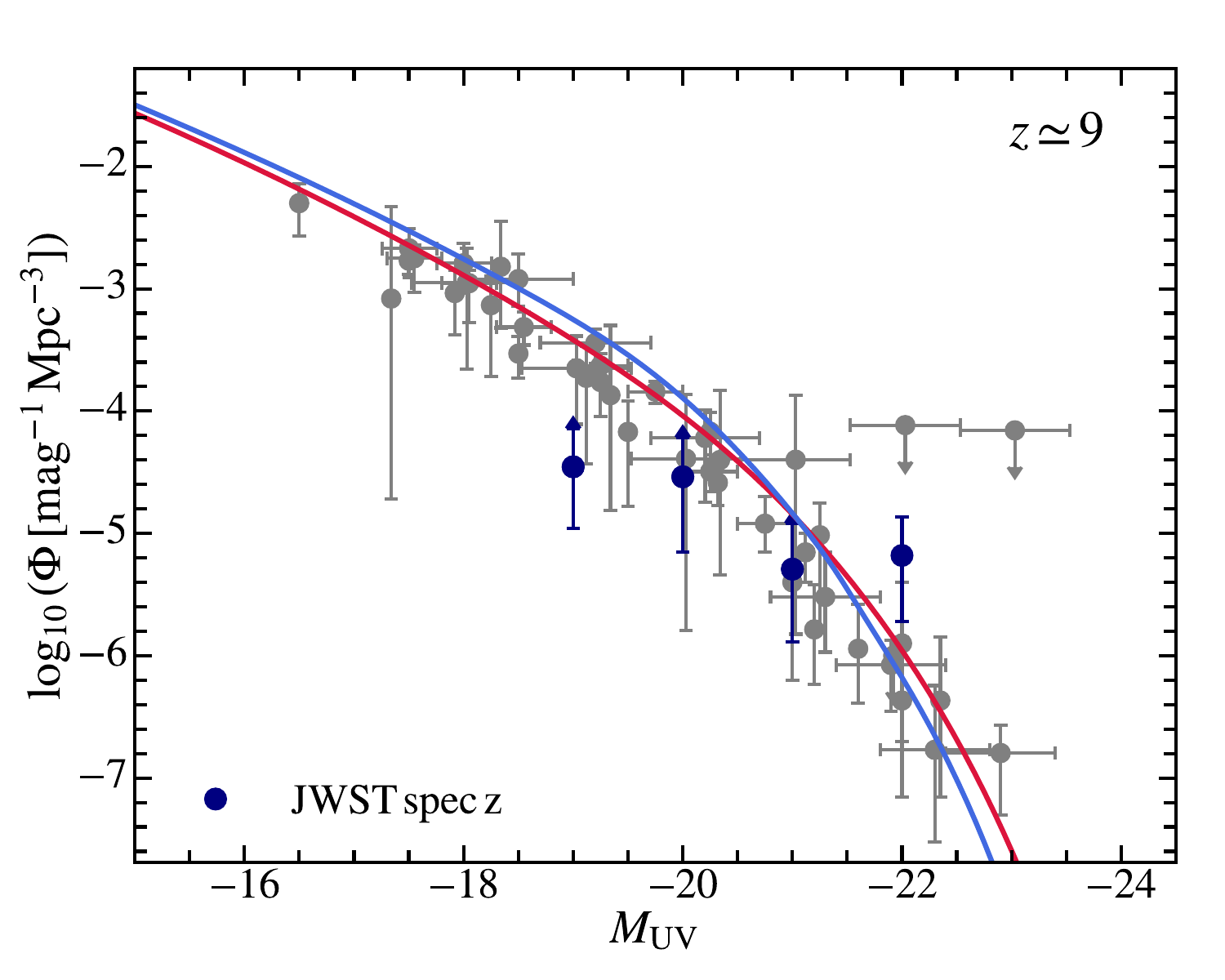}};
    \end{tikzpicture}
    \caption{Galaxy rest-frame UVLFs at $4\lesssim z\lesssim 9$ compared to observational constraints summarized in Section~\ref{sec:uvlf-lowz}. The red (blue) curves show the predictions in $\Lambda$CDM (EDE) assuming a fixed $A$, which is the \textbf{only} parameter we tune here. In the top right panel, we highlight the $z\simeq 6$ case and show the UVLFs with and without dust attenuation. This benchmark model shows remarkable agreement with observations across a wide range of redshifts and UV magnitudes. In particular, the agreement at the faint end is achieved by introducing the halo mass dependence of $\sigma_{\rm UV}$.}
    \label{fig:uvlf-z456}
\end{figure*}

\subsection{Star-formation and UV luminosity}
\label{sec:model-median}

In this section, we describe a \textbf{median} mapping between the observed galaxy UV luminosity and the host halo mass. All the empirical scaling relations we assume should be interpreted as median values. Since the mapping functions we use between $\dot{M}_{\rm halo}$, star-formation rate (SFR), and UV luminosity are monotonic, the median operator is interchangeable with these mapping functions. This ensures that, for example, the mapped UV luminosity from the median $\dot{M}_{\rm halo}$ will be the same as the median UV luminosity mapped from the full $\dot{M}_{\rm halo}$ distribution. 

We parameterize the SFR in dark matter haloes as ${\rm SFR} = \varepsilon_{\ast}\,f_{\rm b}\,\dot{M}_{\rm halo}$, where $f_{\rm b}$ is the universal baryon fraction and $\varepsilon_{\ast}$ is the (halo-scale) SFE. Note that $\varepsilon_{\ast}$ is a differential efficiency in our model while, in some work, the efficiency is defined as the cumulatively formed stellar mass relative to the available halo baryon reservoir. We adopt a redshift-independent double power-law function, 
\begin{equation}
    \varepsilon_{\ast}(M_{\rm halo}) = \dfrac{2 \, \varepsilon_{0}}{ (M_{\rm halo}/M_{0})^{-\alpha} + (M_{\rm halo}/M_{0})^{\beta} }\,,
    \label{eq:sfe}
\end{equation}
where $\varepsilon_{0}$ is the peak SFE at the characteristic mass $M_{0}$ and $\alpha$ and $\beta$ are the low-mass and high-mass end slopes, respectively. The functional form and the redshift-independent ansatz of Equation~\ref{eq:sfe} have been used in previous empirical modeling works~\citep[e.g.][]{Moster2010,Tacchella2018,Harikane2022,Shen2023}. Following \citet{Shen2023}, we adopt $\varepsilon_{0}=0.1$, $M_{0}=10^{12}\msun$, $\alpha=0.6$, $\beta=0.5$ as our default values. The normalization and low-mass slope were chosen to match the median SFR-$M_{\rm halo}$ relation at $z\simeq 7$ from the \textsc{Universe Machine}~\citep{Behroozi2019}. The parameter choices give good agreement with the observed UVLFs and UV luminosity densities at $z\lesssim 9$ as will later be demonstrated in Section~\ref{sec:uvlf-lowz}. 

This model is a basic representation of our knowledge about galaxy formation before the JWST era. In Figure~\ref{fig:sfe}, we compare our model of SFE versus halo mass with other choices in literature, including the observational constraints in \citet{Harikane2022} based on Halo Occupation Distribution, the \textsc{Universe Machine} predictions at $z\simeq 7-12$, the empirical models in \citet{Mason2015,Tacchella2018,Ferrara2023}, results from the FIREbox simulation~\citep{Feldmann2024}, the feedback-free starburst (FFB) scenario at $z\simeq 10$~\citep{Li2023}, and results from the FirstLight simulations~\citep[][which aligns better with the FFB scenario at $z\gtrsim 10$]{Ceverino2024}. Our assumed $\epsilon_{\ast}$-$M_{\rm halo}$ relation is a fair representation of the ``median'' of the models configured before the JWST era. However, we note the substantial uncertainties of the SFE (e.g. $\epsilon_{\ast}$ can vary between $\sim 1-10\%$ at $M_{\rm halo}\sim 10^{10.5}\msun$ among models without entering the FFB regime) and the potential mild increase of SFE at higher redshifts found/assumed in some of these studies~\citep[e.g.][]{Behroozi2019,Ceverino2024}. Degeneracy in matching the UVLF does exist between the low-mass end SFE and other factors~\citep[e.g.][]{Khimey2021,Shen2023,Munoz2023}, in particular UV variability, which will be discussed in the following sections.  

We express the conversion between the SFR and the intrinsic UV-specific luminosity $L_{\nu}(\mathrm{UV})$
(before dust attenuation) as
\begin{equation}
    \label{eq:sfr-luv}
    \mathrm{SFR}\,[\msun\,{\rm yr}^{-1}] = \kappa_{\rm UV} \, L_{\nu}(\mathrm{UV})\,[\erg\,{\rm s}^{-1}\,{\rm Hz}^{-1}]
\end{equation}
with conversion factor $\kappa_{\rm UV} = 0.72 \times 10^{-28}$ as in \citet{Madau2014}, where a \citet{Chabrier2003} IMF is assumed and the (far-)UV wavelength is assumed to be $1500$\AA. 

We empirically model dust attenuation using a combination of the $A_{\rm UV}$-$\beta$ (IRX-$\beta$) and $\beta$-$M_{\rm UV}$ relations. The $M_{\rm UV}$ quoted here is the observed (dust-attenuated) UV magnitude. We adopt the relation $A_{\rm UV} = 4.43 + 1.99 \,\beta$ from \citet{Meurer1999}. We adopt the $\beta$-$M_{\rm UV}$ relation $\beta = -0.17\, M_{\rm UV} - 5.40$ from \citet{Cullen2023} at $8 \lesssim z \lesssim 10$ and the relation from \citet{Bouwens2014} at $z<8$. Motivated by the extremely blue UV slopes of observed galaxies~\citep[e.g.][]{Topping2022, Topping2024, Cullen2024}, we assume no dust attenuation at $z>10$, although uncertainties of dust attenuation in the observed luminous galaxies still exist~\citep[e.g.][]{Bunker2023,Carniani2024,Castellano2024}.

\subsection{UV variability and its dependence on halo mass} 
\label{sec:model-var}

Following \citet{Shen2023}, we calculate galaxy rest-frame UVLF based on the underlying halo mass function as
\begin{equation}
    \dfrac{{\rm d}n}{{\rm d}M_{\rm UV}} = \dfrac{{\rm d}n}{{\rm d} \log_{10}{M_{\rm halo}}} \left| \dfrac{{\rm d} \log_{10}{M_{\rm halo}}}{{\rm d}M_{\rm UV}} \right|,
\end{equation}
and model the stochasticity of UV luminosity at fixed halo mass by convolving the UVLF with a Gaussian kernel of width $\sigma_{\rm UV}$ (in unit of AB magnitude). Effectively, this assumes that the observed UV luminosity has a log-normal distribution with the \textbf{median} value fixed by the scaling relations in Section~\ref{sec:model-median}. $\sigma_{\rm UV}$ effectively captures the scatter of galaxy UV luminosity due to the statistical scatter and time variability of star-formation rates (and dust attenuation) of individual sources. 

\begin{figure*}
    \centering
    \includegraphics[width=0.49\linewidth]{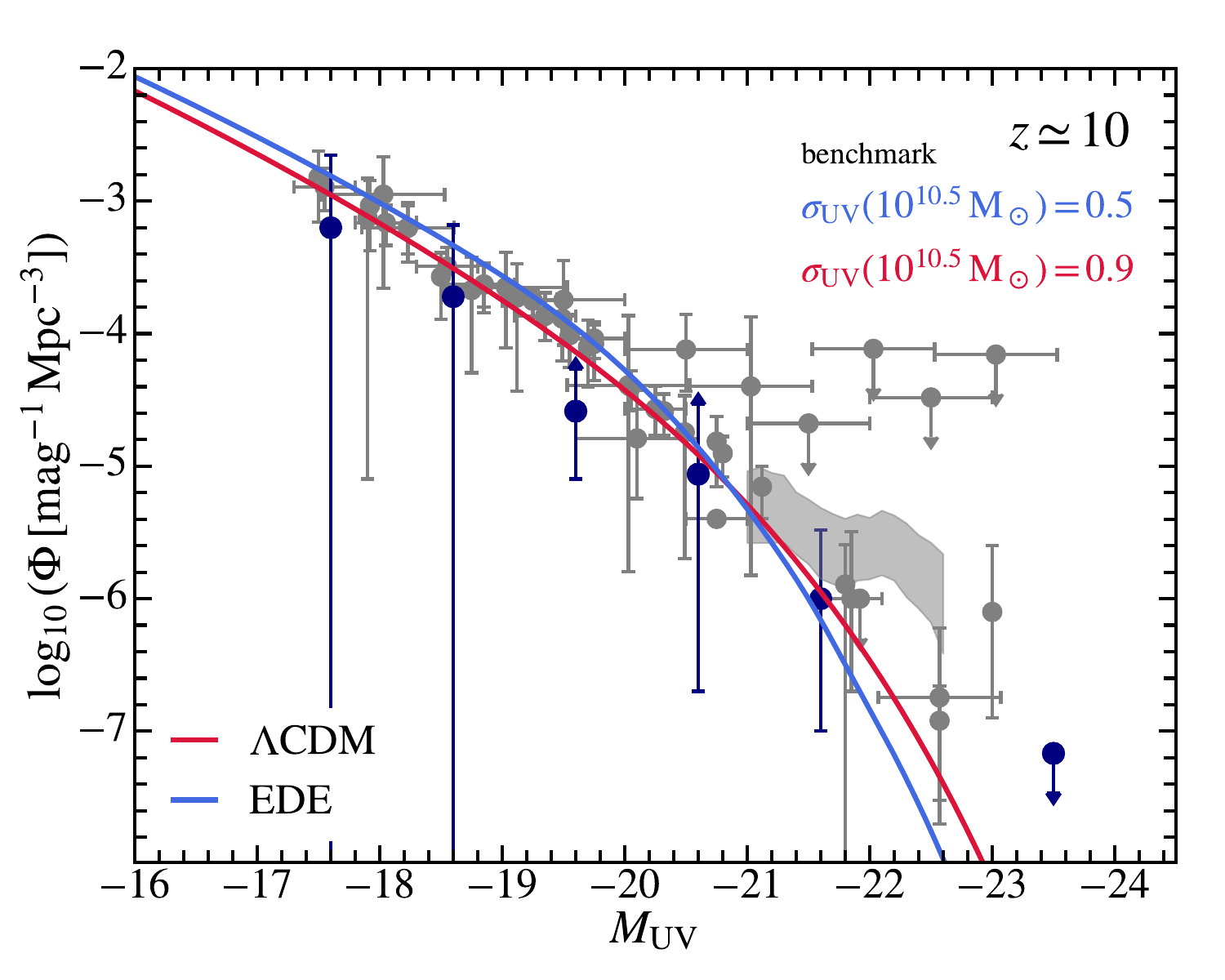}
    \includegraphics[width=0.49\linewidth]{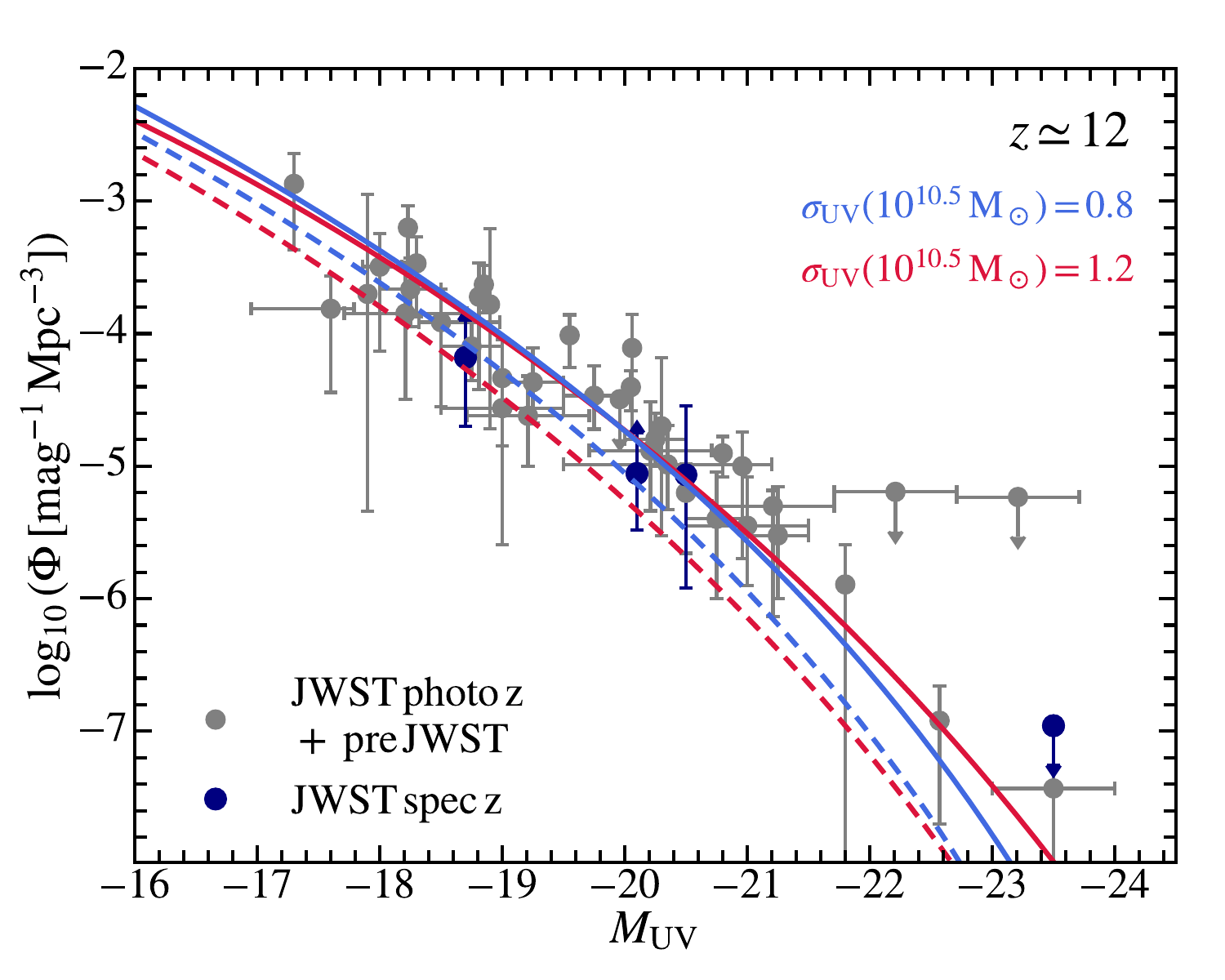}
    \includegraphics[width=0.49\linewidth]{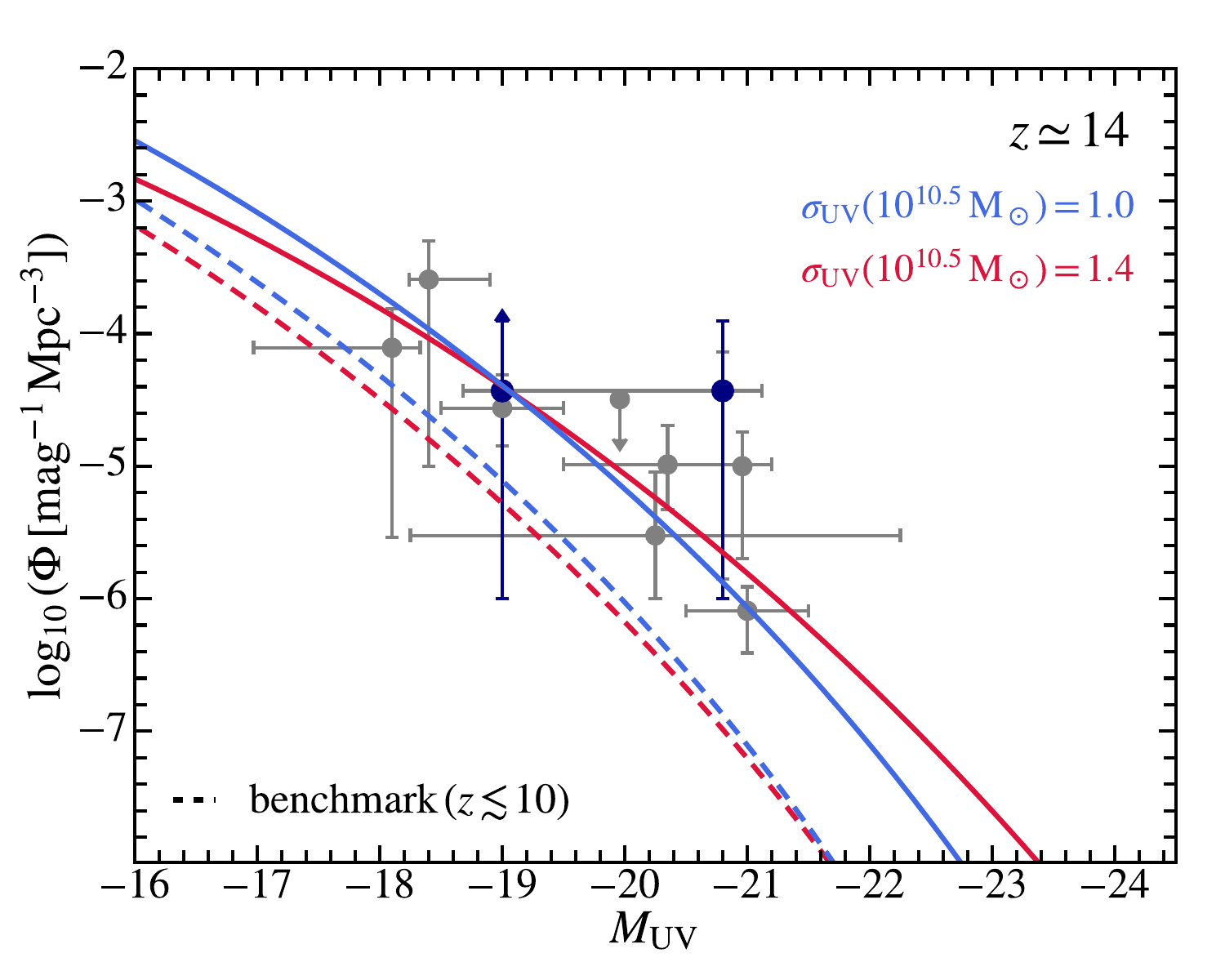}
    \includegraphics[width=0.49\linewidth]{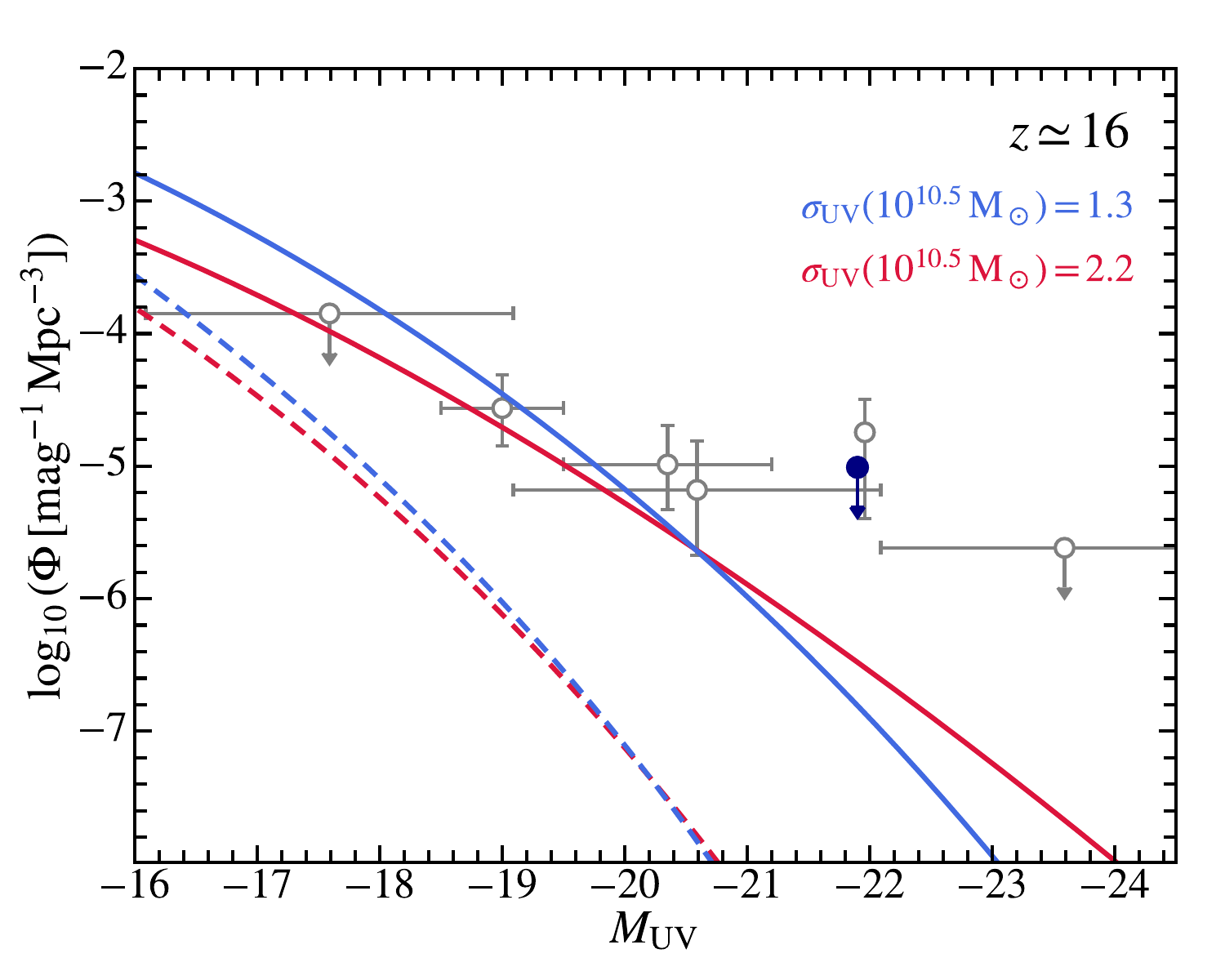}
    \caption{Galaxy rest-frame UVLFs at $z\gtrsim 10$ in $\Lambda$CDM (red) and EDE (blue), compared to observational constraints (the solid points). The observational measurements at $z\simeq 16$ are purely based on photometrically selected galaxies and are shown with open markers. The dashed lines show the predictions of the benchmark model at $z\lesssim 10$. The solid lines show the predictions with the tuning of the normalization of the $\sigma_{\rm UV}$-$M_{\rm halo}$ relation, $A(z)$, although one could also tune the SFE to achieve a similar level of agreement. The suggested values of $\sigma_{\rm UV}$ at $10^{10.5}\msun$ are labelled in each panel. We compare them to the observational constraints summarized in Section~\ref{sec:uvlf-lowz}. The benchmark model in EDE agrees with the spectroscopic constraints at $z\simeq 12$. Although moderately higher $\sigma_{\rm UV}$ is suggested to match the photometric constraints at $z\simeq 12$ and 14, the $\sigma_{\rm UV}$ values are about 0.4 mag smaller than in the $\Lambda$CDM case. Moreover, at $z\simeq 16$, the extreme observational constraints can be reconciled with $\sigma_{\rm UV}$ at $10^{10.5}\msun \simeq 1.3$ mag in EDE while an unphysically high $\sigma_{\rm UV} \simeq 2.2$ mag is required in $\Lambda$CDM.}
    \label{fig:uvlf-z10-16}
\end{figure*}

Motivated by observational results and theoretical model predictions, we propose the following halo mass dependence of $\sigma_{\rm UV}$
\begin{equation}
    \sigma_{\rm UV}(M_{\rm halo}) = {\rm MAX}\left[ A(z) - B\, \log_{10}{(M_{\rm halo}/{\rm M}_\odot)}, \, \sigma_{\rm min} \right],
\end{equation} 
where we choose $B = 0.34$ which follows \citet{Gelli2024}. This has been shown to agree well with the results of cosmological simulations~\citep[e.g.][]{Sun2023,Katz2023}. $A(z)$ is a free parameter that encapsulates the potential redshift dependence of UV variability. $\sigma_{\rm min}=0.2$ mag is the floor of $\sigma_{\rm UV}$ we introduce better match the low-redshift UVLFs at the bright end (see Section~\ref{sec:uvlf-lowz}, although similar agreement can be driven by tuning the massive-end slope $\beta$ of the SFE model). In addition, we assume a maximum value of $\sigma_{\rm max}=2$ mag in low-mass haloes. In Figure~\ref{fig:sigma_vs_mhalo}, we compare this UV variability model at different redshifts with results from various simulations\footnote{We adopt a crude estimate of the host halo mass of the two galaxies presented in \citet{Pallottini2023} based on the stellar-to-halo mass relation in \citet{Behroozi2019}.}
~\citep{Sun2023,Pallottini2023,Katz2023,Feldmann2024}. As will be later introduced in Section~\ref{sec:uvlf-lowz} and \ref{sec:uvlf-highz}, we identify the normalization term $A(z)$ to match the observed UVLFs at different redshifts. We will revisit this figure in Section~\ref{sec:uvlf-highz} when we discuss the UVLFs at $z\gtrsim 10$. 

In this paper, we have assumed a log-normal distribution of observed UV luminosity. However, in practice, a similar phenomenon can be driven by e.g. incorporating a fraction of starbursts with high SFEs. In \citet{Li2023}, a feedback-free starburst (FFB) scenario with $\epsilon^{\rm max}_{\ast}=0.2$ is suggested to explain observational results at $z\gtrsim 10$. Such a scenario can be roughly translated to a feedback-free phase ($\epsilon_{\ast} = 1$) with a duty cycle of 0.2 and otherwise the normal phase ($\epsilon_{\ast} \sim 0.02$). The corresponding standard deviation in $\log_{10}{\epsilon_{\ast}}$ is around $0.68$ dex ($1.7$ mag) above the FFB mass-scale $M_{\rm halo}\sim 10^{10}\msun$ at $z\gtrsim 12$~\citep{Dekel2023}. 

\section{Results}
\label{sec:results}

\begin{figure*}
    \centering
    \includegraphics[width=0.49\linewidth]{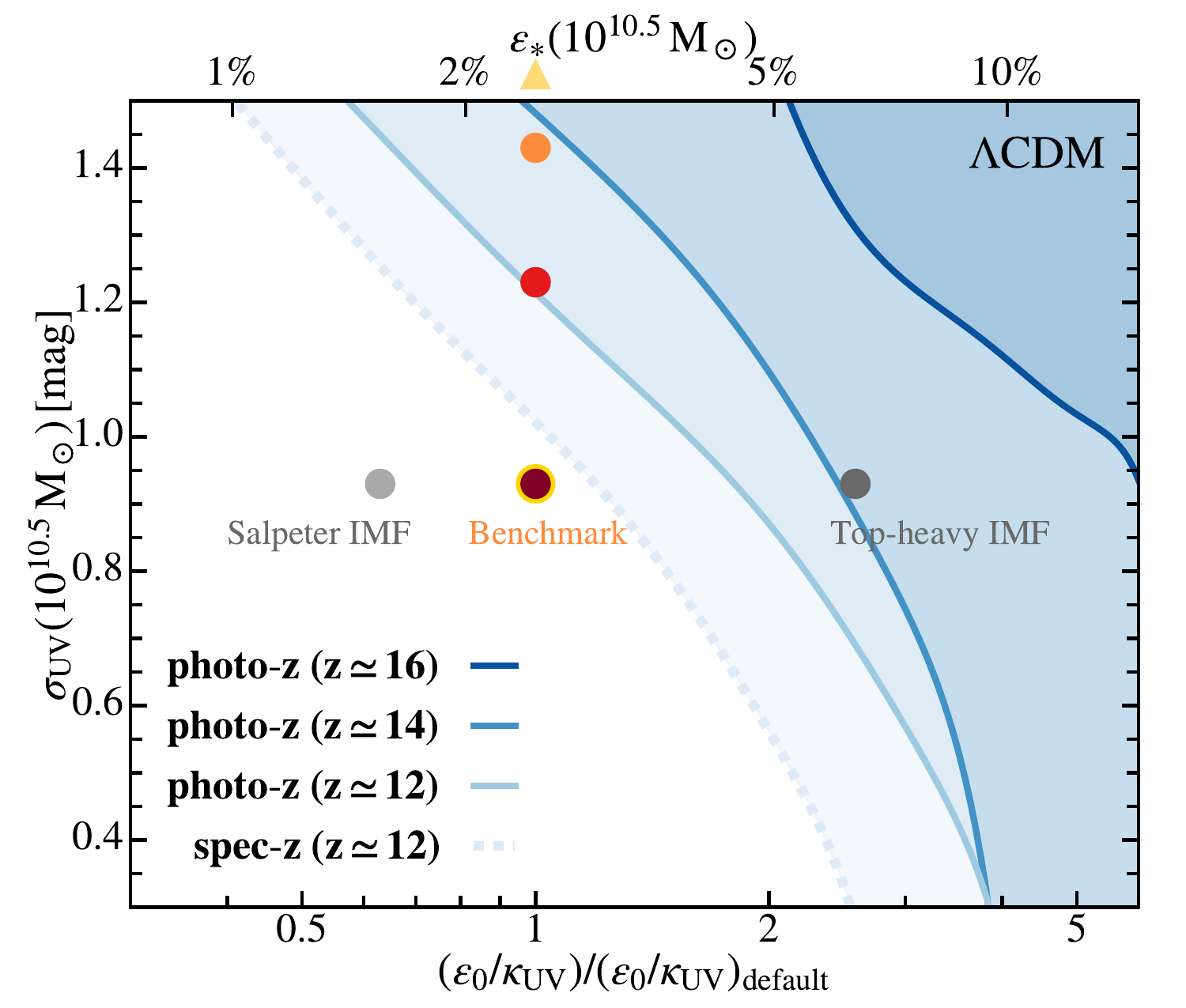}
    \includegraphics[width=0.49\linewidth]{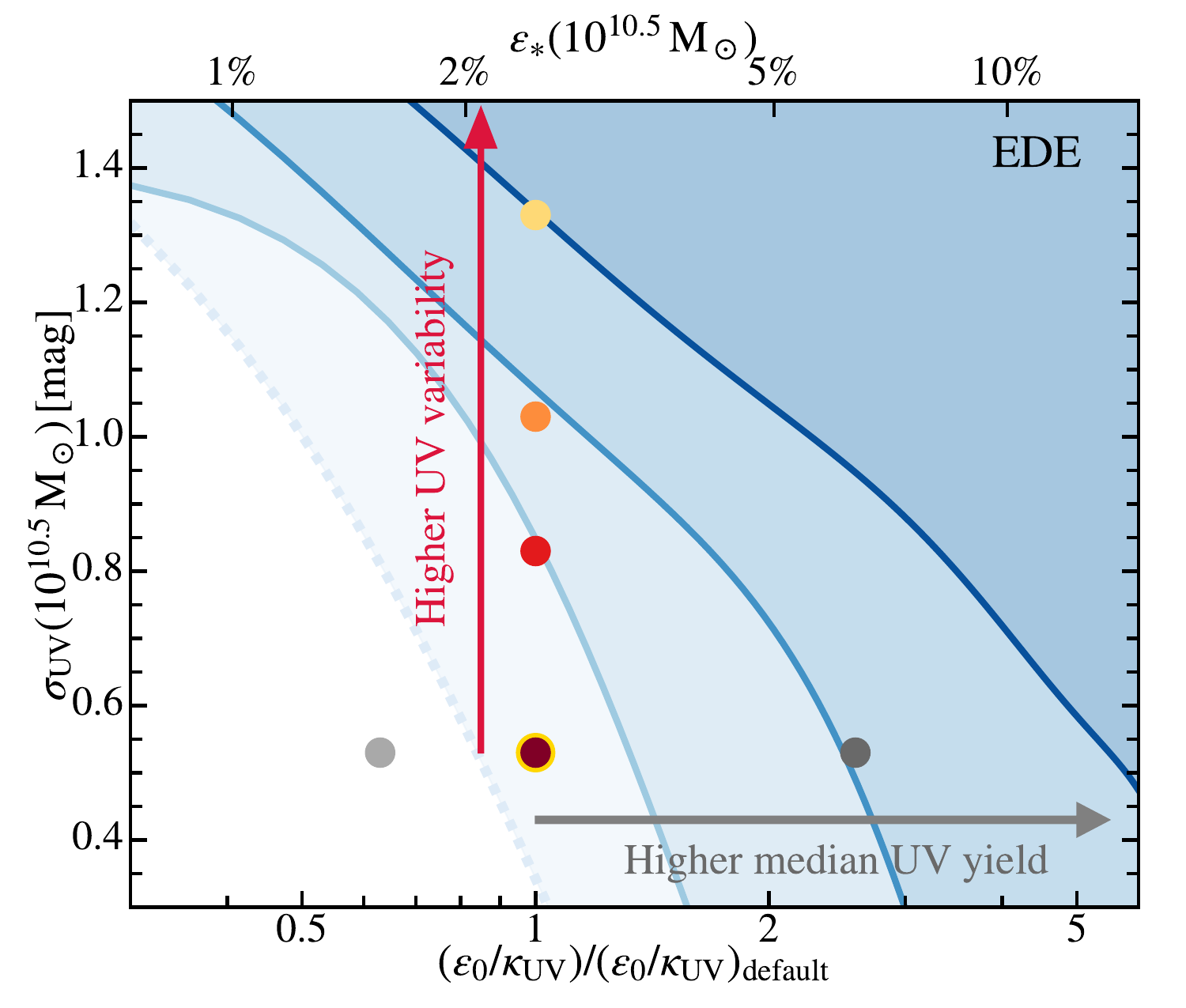}
    \caption{Parameter space of UV variability, characterized by $\sigma_{\rm UV}(10^{10.5}\msun)$, versus median UV radiation yield, represented by $\varepsilon_{0}/\kappa_{\rm UV}$. If the change in median UV radiation yield is purely driven by the change in SFE, the corresponding $\epsilon_{\ast}$ at $M_{\rm halo} = 10^{10.5}\msun$ is labelled. The axis ranges are chosen to roughly cover the $\epsilon_{\ast}$ values assumed in different studies (Figure~\ref{fig:sfe}) and the $\sigma_{\rm UV}$ values found in numeric simulations (Figure~\ref{fig:sigma_vs_mhalo}). The shaded regions show the regions of parameter space consistent with the JWST results mapped by our empirical model. The deeper the color, the more challenging it is to reconcile the constraints. The left (right) panel shows the results in the $\Lambda$CDM (EDE) model. The fiducial model assumes \citet{Chabrier2003} IMF. In the horizontal direction, we show the changes in median UV radiation yield for the \citet{Salpeter1955} IMF and a top-heavy IMF in \citet{Inayoshi2022} with gray circles. In the vertical direction, we show the changes in UV variability required to match different levels of observational constraints with colored circles that match values shown in Figure~\ref{fig:sigma_vs_mhalo} and Figure~\ref{fig:uvlf-z10-16}.}
    \label{fig:parameters}
\end{figure*}

\subsection{Benchmark the model at $4 \lesssim z\lesssim 9$}
\label{sec:uvlf-lowz}

As the first step, we calibrate our model based on the observational constraints of UVLFs at $4 \lesssim z\lesssim 9$. In this paper, we include the HST observations compiled in \citet{Vogelsberger2020} and those from \citet{McLeod2016,Oesch2018,Morishita2018,Stefanon2019,Bowler2020,Bouwens2021}. For JWST constraints, we include constraints based on photometrically selected galaxies from \citet{Castellano2022, Finkelstein2022, Naidu2022, Adams2023b, Bouwens2023a, Bouwens2023b, Donnan2023, Harikane2023, Leetho2023, Morishita2023, Perez2023, Robertson2024, McLeod2024, Donnan2024, Casey2024}. Furthermore, we include constraints based only on spectroscopically-confirmed galaxies~\citep[][see the references therein]{Harikane2024-spec,Harikane2024b-spec}.

We adopt the redshift-independent but halo mass-dependent SFE described in Section~\ref{sec:model-median} and only allow the normalization term $A(z)$ of UV variability to vary (as the variability is less constrained compared to SFE in literature). Notably, we find that a redshift-independent choice of $A = 4.5$ ($A = 4.1$) in $\Lambda$CDM (EDE) leads to UVLFs that are consistent remarkably with observations across the entire redshift range. In Figure~\ref{fig:uvlf-z456}, we show the results of the calibrated models compared to the observed UVLF at $4\lesssim z \lesssim 9$. Compared to \citet{Shen2023}, the halo mass dependence of $\sigma_{\rm UV}$ helps us realize a better agreement with the faint-end of UVLF across redshifts. Similar results have recently been demonstrated in \citet{Gelli2024}. These preferred $\sigma_{\rm UV}(M_{\rm halo})$ relations at $z\lesssim 10$ are also shown in Figure~\ref{fig:sigma_vs_mhalo}. This calibrated model will be referred to as the benchmark model in the following analysis.

\subsection{UVLF at $z\gtrsim 10$}
\label{sec:uvlf-highz}

We now explore the model variations necessary at $z\gtrsim 10$ to reconcile observations with theoretical model predictions in $\Lambda$CDM versus EDE. Due to the degeneracy between the SFE in the low-mass end and UV variability on UVLF predictions, variation in either direction has the potential to resolve the UVLF tension (see also \citealt{Munoz2023}). In this Section, we focus on varying the UV variability by adjusting the normalization $A(z)$ and defer the discussion on the alternative SFE solution to Figure~\ref{fig:parameters}. Since the slope of the $\sigma_{\rm UV}$-$M_{\rm halo}$ relation is fixed, we will denote the change to $A(z)$ with $\sigma_{\rm UV}$ at a characteristic halo mass $10^{10.5}\msun$. In the benchmark model, $\sigma_{\rm UV}(10^{10.5}\msun)=0.9$ and 0.5 mag in the $\Lambda$CDM and EDE cosmologies, respectively.

Figure~\ref{fig:uvlf-z10-16} shows the UVLFs obtained from our model at $z\gtrsim 10$. At $z\simeq 10$, predictions from the benchmark models in both $\Lambda$CDM and EDE continue to match observations without further tuning. At $z\simeq 12$, to match the spectroscopic constraints, a small $\lesssim 0.1$ mag enhancement of UV variability is sufficient for the $\Lambda$CDM. Meanwhile, the benchmark model in EDE works without any change to $\sigma_{\rm UV}(10^{10.5}\msun)$. Similar findings have been discussed in \citet{Gelli2024}, where the halo mass dependence of UV variability can reduce the need for model variations at $z\simeq 12$. To match the photometric constraints at the same redshift, $\sigma_{\rm UV}(10^{10.5}\msun)$ needs to be enhanced to $\simeq 1.2$ (0.8) mag to match observational constraints in $\Lambda$CDM (EDE). However, the discrepancies between observations and the predictions of benchmark models are relatively small ($\lesssim 0.3\,{\rm dex}$) and fall within the uncertainties from halo mass function calculations~\citep[e.g.][]{Yung2024} and cosmic variances~\citep[e.g.][]{Yung2023,KJ2024}.

At $z\simeq 14$, $\sigma_{\rm UV}(10^{10.5}\msun) \simeq 1.4$ (1.0) mag is preferred in $\Lambda$CDM (EDE), although the benchmark model predictions are still within the 1-$\sigma$ error bar of observational constraints from e.g. \citet{Donnan2024}. Two galaxy candidates at $z\simeq 14$ have recently been spectroscopically confirmed~\citep{Carniani2024} and show signatures of a strong recent starburst in MIRI band~\citep{Helton2024}. At $z\simeq 16$, in $\Lambda$CDM, we find that a simple adjustment of $A(z)$ is no longer feasible to reconcile observations and we have to increase $\sigma_{\rm UV}$ to $\simeq 2.2$ mag, ignoring the $\sigma_{\rm max}$ we imposed. The model becomes equivalent to the constant $\sigma_{\rm UV}$ model explored in \citet{Shen2023}~\footnote{The small difference in $\sigma_{\rm UV}$ inferred is mainly caused by changing the IMF to \citet{Chabrier2003} in this work from \citet{Salpeter1955} in \citet{Shen2023}.}. We note that the UV variability in the massive end will not impact the UVLF in the observed range~\footnote{We obtain this mass range by manually setting UV variability above some mass threshold to the benchmark model values and check if the resulting UVLF is still consistent with observations.}, as indicated by the dashed lines in Figure~\ref{fig:sigma_vs_mhalo}. Nevertheless, such a high $\sigma_{\rm UV}$ exceeds the maximum value found in cosmological hydrodynamical simulations (see Figure~\ref{fig:sigma_vs_mhalo}). As discussed in \citet{Kravtsov2024}, it could lead to instantaneous SFE exceeding unity when modelling the bursty star-formation history of galaxies. On the contrary, in the EDE model, $\sigma_{\rm UV}(10^{10.5}\msun)$ only needs to be enhanced moderately to $\simeq 1.3$ mag to reconcile observations with theoretical model predictions. This significantly eases the level of model adjustment to account for these extreme constraints. We summarize these suggested $\sigma_{\rm UV}(M_{\rm halo})$ at different redshifts in Figure~\ref{fig:sigma_vs_mhalo}. In $\Lambda$CDM, the implied $\sigma_{\rm UV}$ values at $z\simeq 14$ are close to the maximum value found in existing cosmological hydrodynamical simulations and the $\sigma_{\rm UV}$ values at $z\simeq 16$ exceed substantially this maximum. However, in the EDE cosmology, the suggested $\sigma_{\rm UV}$ values are safely among the scatters of simulation results.

\subsection{Explore the model parameter space}

To illustrate various model variations to reconcile JWST results with theoretical models, Figure~\ref{fig:parameters} examines the parameter space of UV variability and the median UV radiation yield in the $\Lambda$CDM and EDE cosmology. For results based on JWST spectroscopy, we adopt the lower limit estimated in \citet{Harikane2024-spec,Harikane2024b-spec} at $z\simeq 12$. For photometric constraints, we consider the model to be acceptable when $\log_{10}{\Phi}(M_{\rm UV}=-20)>-4.8$ at $z\simeq 12$, and $\log_{10}{\Phi}(M_{\rm UV}=-19)>-4.4$ and $-4.5$ at $z=$14 and 16. These are fairly rough estimates (i.e. without a quantitative measure of fitting residuals), but we have explicitly examined the UVLF predictions around these threshold values and find that they capture the goodness of fit reasonably well. We scan the parameter space by modifying UV variability ($A(z)$ or equivalently $\sigma_{\rm UV}(10^{10.5}\msun)$) and the normalization of the SFE in our model, and identify the regime where theoretically predicted UV bright galaxy abundance exceeds the observed values. We also show the impact of IMF variations as in \citet{Shen2023}, where a Salpeter IMF leads to about $60\%$ increase in $\kappa_{\rm UV}$~\citep{Madau2014} and an extremely top-heavy IMF~\citep[e.g.][]{Inayoshi2022} leads to $\kappa_{\rm UV}$ dropping by $55\%$. 

In $\Lambda$CDM, the benchmark model is only slightly off from the spectroscopic constraints of JWST. The required model variations are much smaller compared to \citet{Shen2023}. This is mainly due to the change to \citet{Chabrier2003} IMF and the indirect enhancement of UV variabilities from the halo mass dependence. Galaxies with the same UV luminosity are hosted by lower mass haloes at higher redshifts and thus exhibit higher $\sigma_{\rm UV}$ even in the absence of redshift-dependence of $\sigma_{\rm UV}$. In the EDE cosmology, the out-of-the-box prediction at $z\simeq 12$ is fully consistent with the spectroscopic constraints. As also demonstrated in Figure~\ref{fig:uvlf-z10-16}, the UV variability needs to be increased to match photometric constraints at $z\simeq 12$ and 14. The required values of $\sigma_{\rm UV}(10^{10.5}\msun)$ in EDE are significantly smaller than those in $\Lambda$CDM and are more consistent with predictions from numeric simulations. Alternatively, one could also reconcile the observations by boosting the SFE or incorporating a top-heavy IMF to increase the light-to-mass ratio. Along this orthogonal direction, agreement with observations is also easier to achieve in the EDE cosmology compared to the  $\Lambda$CDM case. A small factor of $\sim 1.5$ (2.5) boost in UV photon yield is necessary for the $z\simeq 12$ (14) photometric constraints, which corresponds to $\epsilon_{\ast}(10^{10.5}\msun) \simeq 3\%$ ($6\%$). These are feasible given the uncertainties of SFE in low-mass haloes as shown in Figure~\ref{fig:sfe}.

Qualitative differences show up when confronting the $z\simeq 16$ constraints. In $\Lambda$CDM, an extremely high, constant $\sigma_{\rm UV} \simeq 2.2$ mag is required, which would imply a strong non-linear transition of UV variability at $z\gtrsim 14$. In the EDE cosmology, it becomes feasible with a suggested $\sigma_{\rm UV}(10^{10.5}\msun)\simeq 1.3$ mag. One could also achieve this by boosting the SFE by a factor of $\sim 5$ and the implied $\epsilon_{\ast}$ at $\sim 10^{10.5}\msun$ will be $\sim 10\%$, which is below the maximum value found in literature and could be easily reached in the FFB scenario~\citep{Li2023}. The introduction of EDE eases the level of tuning to the galaxy formation model to reconcile these extreme JWST results. However, we are cautious in making strong conclusions about $\Lambda$CDM based on the comparison at this redshift, since the constraints here are largely based on a handful of photometrically-selected galaxies~\citep[e.g.][]{Bouwens2023a,Harikane2023} and remain uncertain.

\begin{figure}
    \centering
    \includegraphics[width=\linewidth]{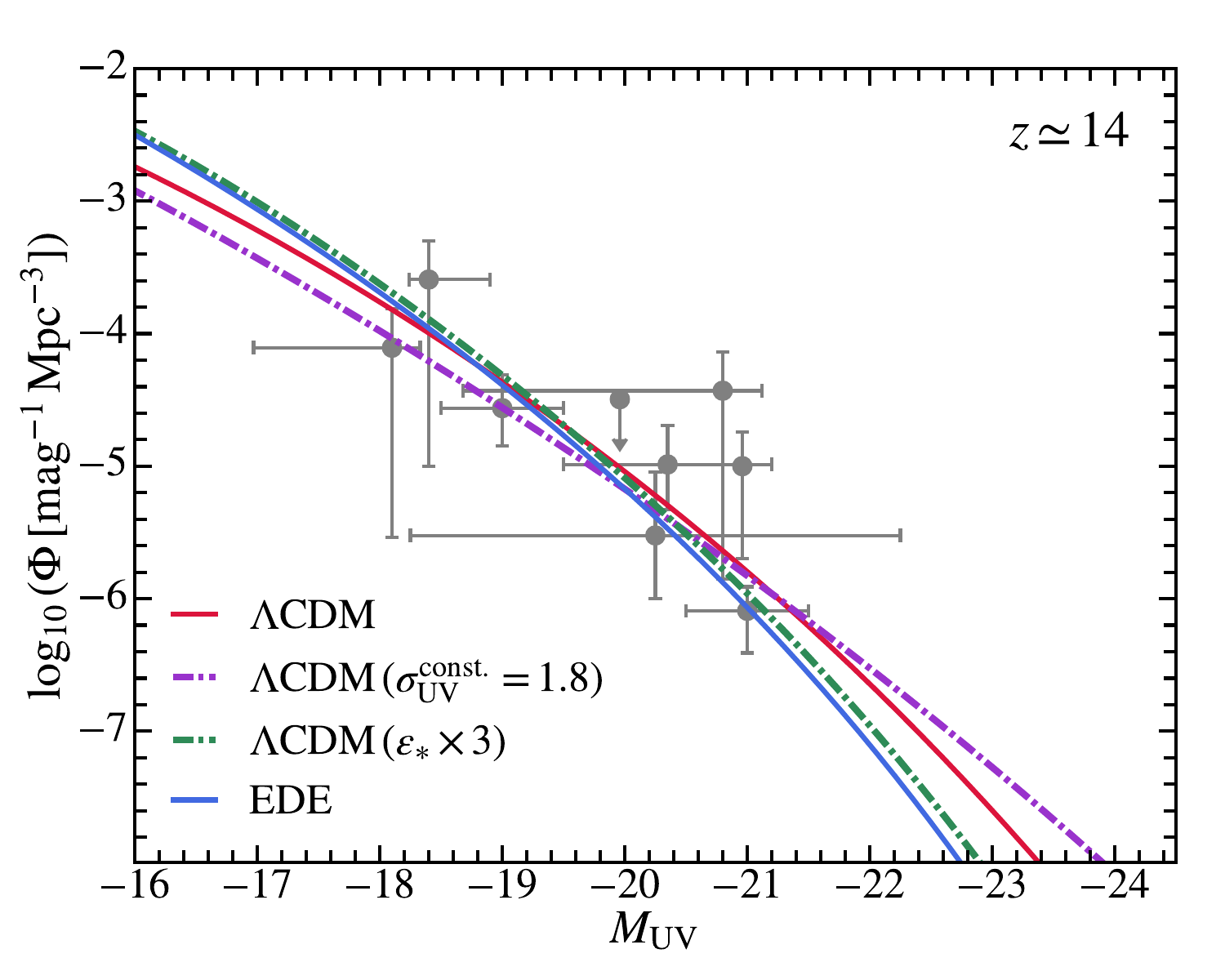}
    \caption{Galaxy UVLF at $z\simeq 14$ in the following four scenarios: (1) the constant UV variability model considered in \citet{Shen2023}; (2) boosting the SFE by a factor of three while assuming the benchmark $\sigma_{\rm UV}(M_{\rm halo})$ in $\Lambda$CDM; (3) adopt appropriate modifications to the normalization of $\sigma_{\rm UV}(M_{\rm halo})$ in $\Lambda$CDM and (4) EDE, respectively. The observed UVLF at $z\simeq 14$ are equally well explained by these adjustments to the galaxy formation model.}
    \label{fig:uvlf-variations}
\end{figure}

\begin{figure}
    \centering
    \includegraphics[width=\linewidth]{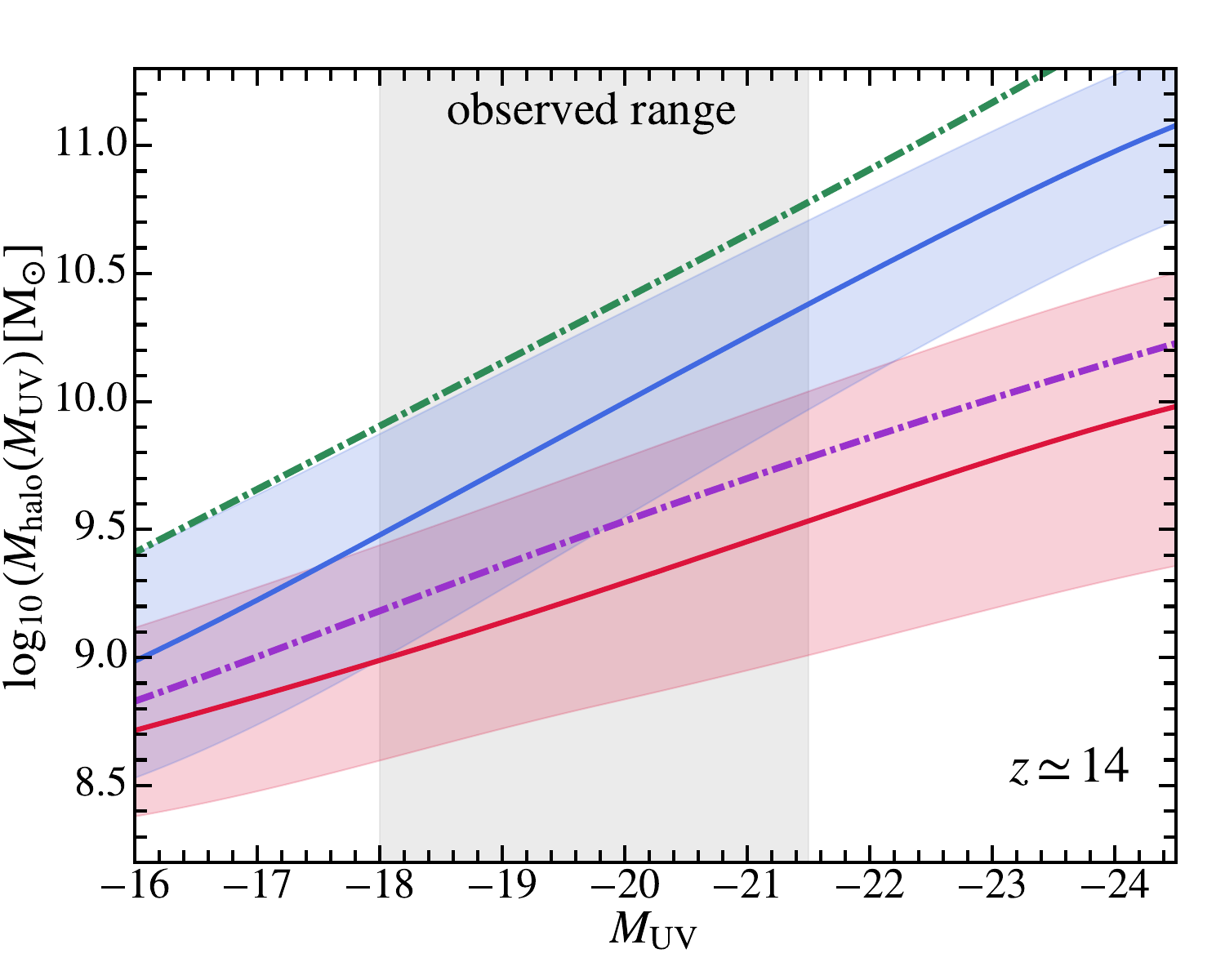}
    \includegraphics[width=1.002\linewidth]{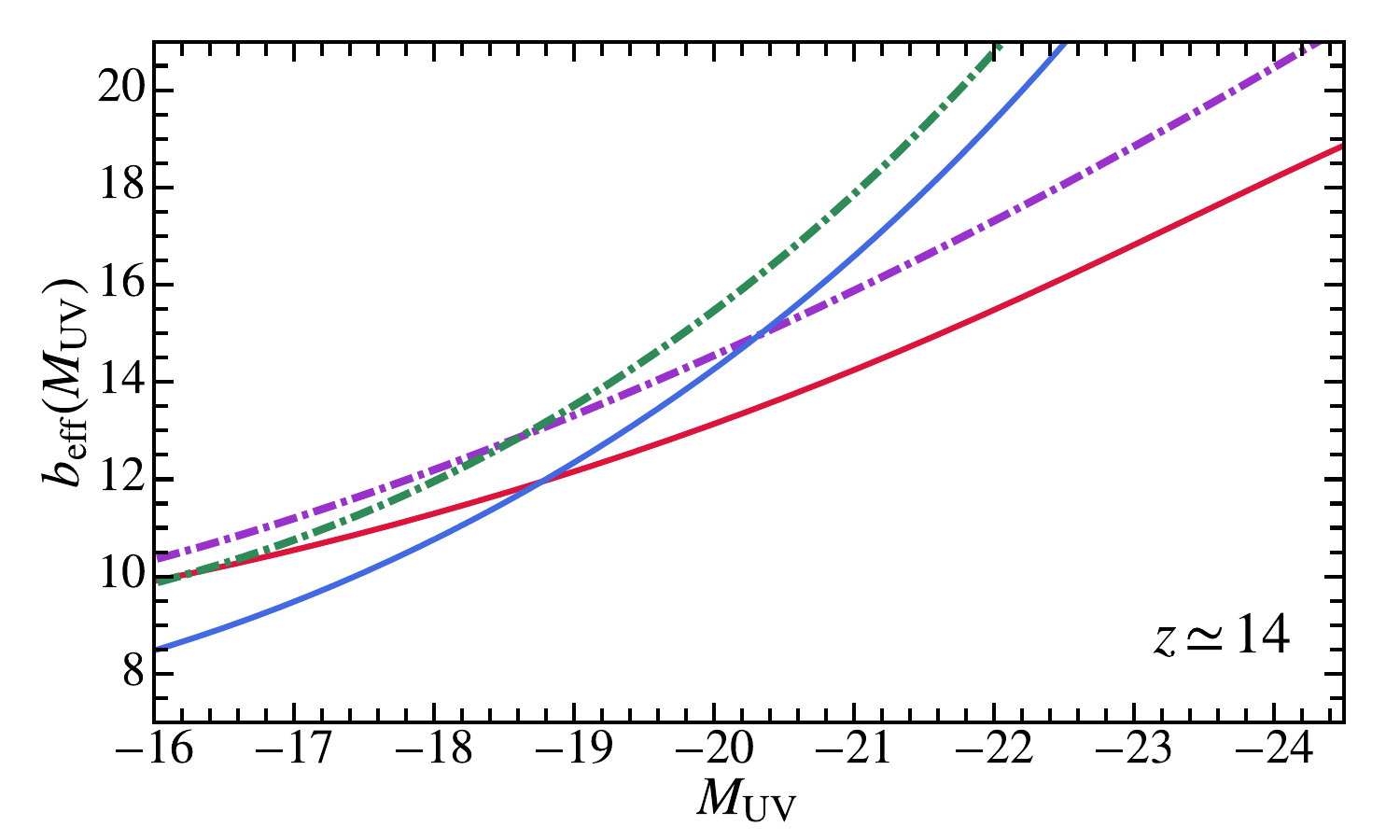}
    \caption{{\it Top}: Median halo mass at a given observed UV magnitude, $M_{\rm UV}$, at $z\simeq 14$. We consider the same four scenarios as shown in Figure~\ref{fig:uvlf-variations} and adopt the same labelling. The gray-shaded region shows the typical luminosity range of observed galaxies at this redshift. {\it Bottom}: Effective bias of galaxies as a function of $M_{\rm UV}$. Due to the lower host halo mass in model variations with larger UV variability, the bias of bright galaxies is significantly reduced.}
    \label{fig:clustering}
\end{figure}

\subsection{Host haloes of UV-bright galaxies at cosmic dawn}

In Figure~\ref{fig:uvlf-variations}, we show the UVLFs at $z\simeq 14$ in four different scenarios that could equally well match observational constraints. This includes the model with constant $\sigma_{\rm UV}$ as explored in \citet{Shen2023}, the approach of boosting the SFE by three times with respect to the benchmark model at low redshifts (see similar values found in e.g. \citealt{Yung2023}), and the approach of adjusting the normalizations of $\sigma_{\rm UV}(M_{\rm halo})$ in the $\Lambda$CDM and EDE cosmologies as discussed above. However, due to different levels of $\sigma_{\rm UV}$ adopted, the host halo mass of observed UV-bright galaxies can be rather different in these scenarios. In the top panel of Figure~\ref{fig:clustering}, we show the median host halo mass of galaxies at a given $M_{\rm UV}$ at $z\simeq 14$. Assuming only varying UV variability, the typical halo mass of observed galaxies in $\Lambda$CDM is around $10^{9}\msun$ at $z\simeq 14$, the mass scale of current-day ultra-faint dwarfs. The descendants of these haloes will end up with $\sim 10^{13}$ - $10^{14}\msun$ when evolving to $z=0$~\citep[e.g.][]{Lu2024}. However, in the EDE cosmology, due to the much smaller $\sigma_{\rm UV}$ required, the host haloes are about $0.5$ - $1$ dex heavier than in $\Lambda$CDM. It is even more of the case if one attempts to solve the tension by only adjusting SFE. We find that the host halo mass in this scenario will be around $10^{10.5}\msun$.

The different host halo masses can lead to different clustering powers of observed UV-bright galaxies. Following e.g. \citet{Munoz2023,Gelli2024}, we calculate the number density-normalized, effective bias of galaxies at fixed UV luminosity as
\begin{equation}
    b_{\rm eff}(M_{\rm UV}) = \dfrac{1}{\Phi(M_{\rm UV})}\,\int {\rm d}M_{\rm halo}\,\dfrac{{\rm d}n}{{\rm d}M_{\rm halo}}\,b(M_{\rm halo})\,P(M_{\rm halo}|M_{\rm UV}),
\end{equation}
where $\Phi(M_{\rm UV})$ is the UVLF, $P(M_{\rm halo}|M_{\rm UV})$ is the conditional probability distribution of halo mass at a given $M_{\rm UV}$, $b(M_{\rm halo})$ is the halo bias calculated using the \citet{Tinker2010} model as implemented in the \textsc{Colossus} package~\citep{Diemer2018}. In the bottom row of Figure~\ref{fig:clustering}, we show $b_{\rm eff}$ versus the observed $M_{\rm UV}$. Models with higher UV variability predict lower $b_{\rm eff}$ in general. The differences show up primarily at the bright end. It is worth noting that even assuming the same galaxy-halo mapping, the bias in the EDE cosmology will still be smaller than that in $\Lambda$CDM due to the larger $H_0$ value. This makes the galaxy bias difference between EDE and $\Lambda$CDM smaller than naively inferred from the $M_{\rm UV}$-$M_{\rm halo}$ relation. Measurements of the environment of these UV-bright galaxies may help distinguish these degenerate scenarios. Potential avenues include measurements of reionization bubble sizes~\citep[e.g.][and see a recent theoretical study by \citealt{Neyer2024}]{Hsiao2023,Umeda2023,Nadler2023}, simple neighbor searches~\citep[e.g.][]{Tacchella2023}, and future galaxy clustering measurements with JWST and Roman.

\begin{figure*}
    \centering
    \includegraphics[width=0.33\linewidth, trim={0.5cm 0 0.5cm 0}]{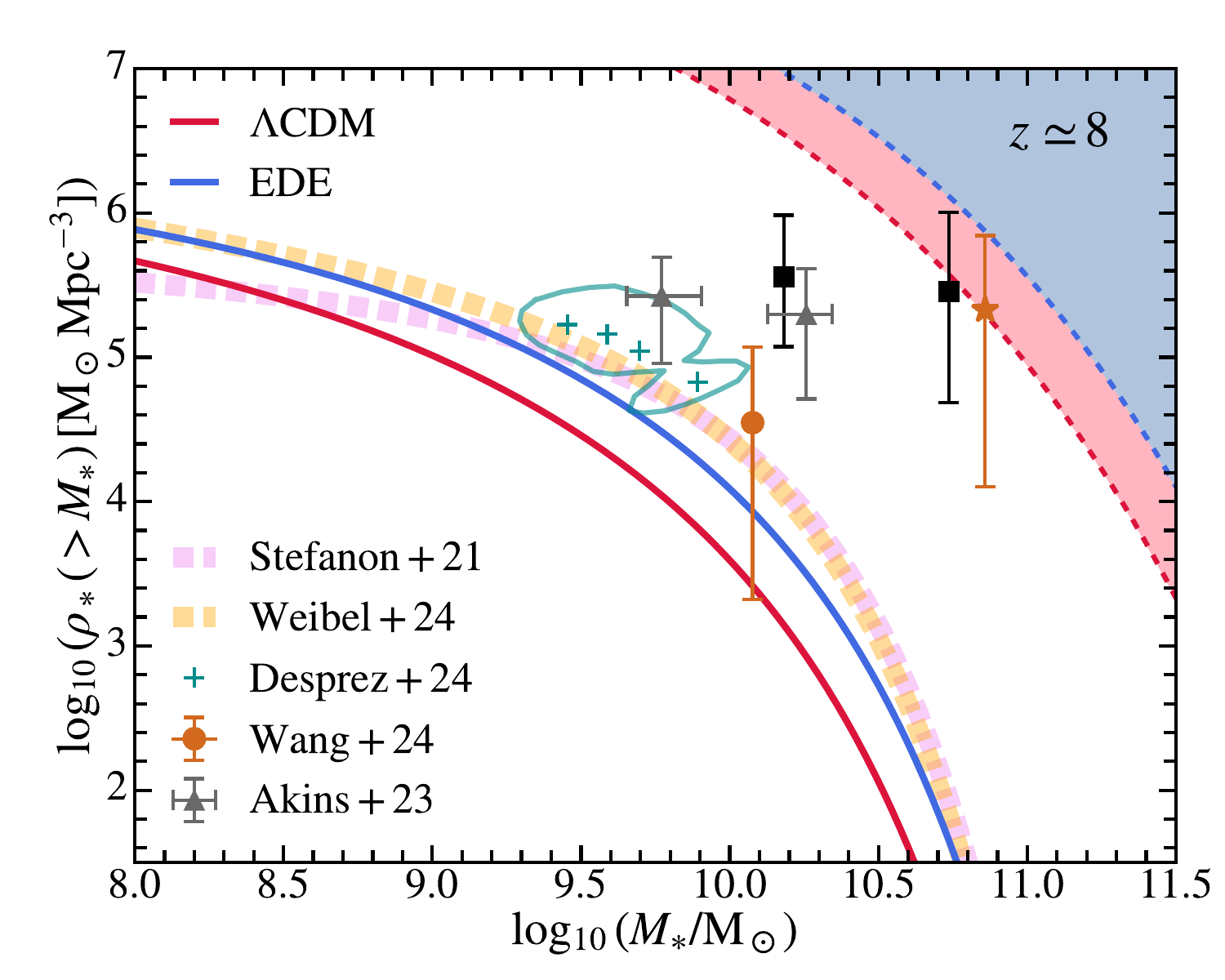}
    \includegraphics[width=0.33\linewidth, trim={0.5cm 0 0.5cm 0}]{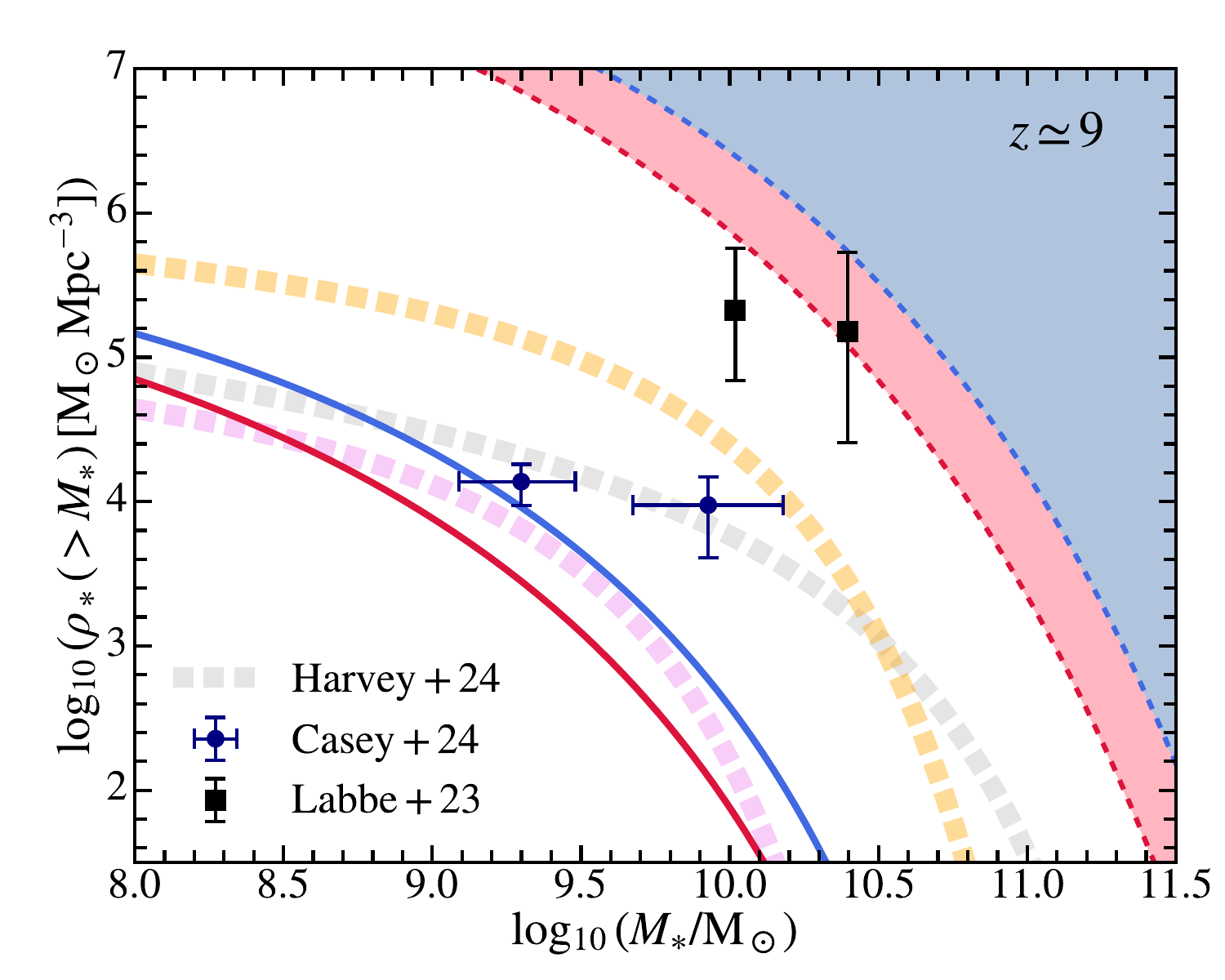}
    \includegraphics[width=0.33\linewidth, trim={0.5cm 0 0.5cm 0}]{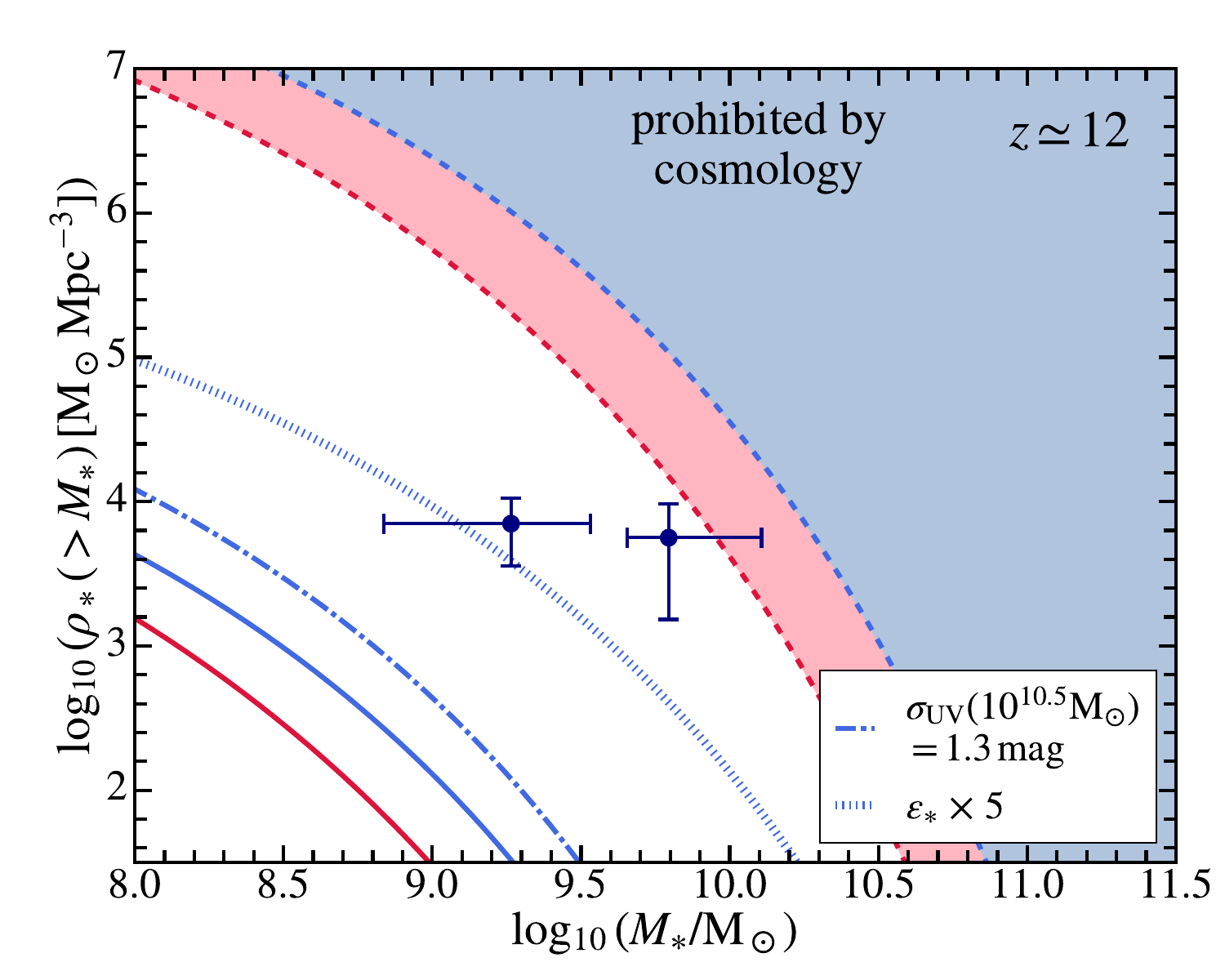}
    \caption{Cosmic stellar mass density for galaxies with stellar mass above $M_{\ast}$ at $z\simeq 8$, 9, and 12. The dashed lines show the upper limit derived by converting all the baryons in the Universe to stars in the assumed cosmology. The solid lines show predictions from our benchmark model at $z\simeq 8,$ 9, and the slightly tuned model at $z\simeq 12$ which match the UVLF constraints in Figure~\ref{fig:uvlf-z10-16}. The dotted and dotted-dashed lines show two alternative scenarios in EDE when (1) $\sigma_{\rm UV}(10^{10.5}\msun)$ is further increased to $1.3$ mag (the value needed for the $z\simeq 16$ UVLF constraints) or (2) SFE is enhanced by a factor of five. We compare them to the observational constraints from JWST reported in \citet{Labbe2023,Akins2023,Casey2024}, and \citet{Wang2024-RUBIES} (``medium'' estimate as the circle, ``maximum'' as the star). \citet{Desprez2024} revisited the CEERS sample used in \citet{Labbe2023} and derived alternative constraints when the single most massive outlier was removed. We also show estimates by integrating the Schechter function fits of stellar mass functions from \citet{Stefanon2021,Weibel2024,Harvey2024}. IMF-related corrections on stellar masses are made for all these data. The most extreme observational constraints from JWST, which are in conflict with $\Lambda$CDM, can be safely accommodated in the EDE cosmology. The benchmark model predictions at $z\simeq 8$ and 9 marginally agree with some less stringent observational constraints. At $z\simeq 12$, a boosted SFE can better explain the large cosmic stellar mass densities compared to enhanced UV variabilities.}
    \label{fig:rhostar}
\end{figure*}

\subsection{Implications for the overly-massive galaxies}

Another natural consequence of enhanced halo abundance in EDE is the increased number density of massive galaxies. Early JWST observations have revealed several overly massive galaxy candidates exceeding expectations in the standard galaxy formation model given the field of view of observations~\citep[e.g.][]{Labbe2023,Xiao2023,Casey2024}. Some even require converting all the baryons in the Universe to stars to produce, thus challenging the $\Lambda$CDM model~\citep[e.g.][]{BK2023,Lovell2023}. While the stellar mass estimates of these galaxies are uncertain~\citep[e.g.][]{Endsley2022,Chworowsky2023,Narayanan2024,Wang2024} and even their ``galaxy or AGN'' identification is subject to debates~\citep[e.g.][]{Kocevski2023,Desprez2024,Wang2024}, our exploration here delves into how this tension would manifest within the framework of EDE cosmology.

We derive the stellar-to-halo mass relation of galaxies by integrating the SFE in Equation~\ref{eq:sfe}
\begin{align}
    M_{\ast}(M_{\rm halo}) = & (1-R)\, \bigg[\epsilon_{\ast}(M_{\rm min})\,M_{\rm min} \\
    & + \int^{M_{\rm halo}}_{\rm M_{\rm min}} {\rm d}M^\prime_{\rm halo}\,f_{\rm b}\,\epsilon_{\ast}(M^\prime_{\rm halo})\,\mathcal{F}(\sigma_{\rm sf})\bigg], \nonumber
\end{align}
where $M_{\rm min}=10^8 \msun$ is the minimum halo mass we start from (the SFE below which is assumed to be a constant) which does not affect our result, $R$ is the mass return fraction of stars taken to be $0.1$ for the young and low-metallicity stellar populations at high redshifts~\citep[e.g.][]{Hopkins2018,Hopkins2023-fire3,Feldmann2024}, $\mathcal{F}(\sigma_{\rm sf})$ is the factor considering the difference between mean SFR and median SFR with the logarithm scatter $\sigma_{\rm sfr}$ in unit of dex
\begin{equation}
    \mathcal{F}(\sigma_{\rm sfr}) = \exp{\left(\ln{10}\,\sigma_{\rm sfr}^2/2\right)}.
\end{equation}
If the UV variability is dominated by the burstiness of star-formation, we have $\sigma_{\rm sfr} = \sigma_{\rm UV}/2.5$. We include this factor when calculating the integrated stellar mass density, but remove it when calculating e.g. median stellar mass given halo mass. We have verified that this approach gives consistent stellar-to-halo mass ratios with abundance-matching results~\citep[e.g.][]{Behroozi2013,Behroozi2019,RP2017}. We obtain the cumulative cosmic stellar mass density above $M_{\ast}$ as
\begin{equation}
    \rho_{\ast}(>M_{\ast}) = \int^{\infty}_{M_{\rm halo}(M_{\ast})}\,{\rm d}M^\prime_{\rm halo}\,\dfrac{{\rm d}n}{{\rm d}M^\prime_{\rm halo}}\,M_{\ast}(M^\prime_{\rm halo}),
    \label{eq:rhostar}
\end{equation}
where we effectively assume the scatter in galaxy stellar mass is negligible after aggregating many cycles of stochastic star-formation. The maximum stellar mass density allowed by cosmology is
\begin{equation}
    \rho^{\rm max}_{\ast}(>M_{\ast}) = \int^{\infty}_{M_{\rm halo}(M_{\ast})}\, {\rm d}M^\prime_{\rm halo}\,\dfrac{{\rm d}n}{{\rm d}M^\prime_{\rm halo}}\,f_{\rm b}\,M^\prime_{\rm halo}.
\end{equation}

In Figure~\ref{fig:rhostar}, we compare the stellar mass density in $\Lambda$CDM and EDE with observational constraints from JWST. We show the model predictions at $z\simeq 8$, 9, and 12, which corresponds to the constraints from CEERS~\citep{Labbe2023}, COSMOS-Web~\citep{Casey2024}, RUBIES~\citep{Wang2024-RUBIES}, and \citet{Akins2023}. For reference, we also show estimates by integrating the Schechter function fits of stellar mass functions from \citet{Stefanon2021,Weibel2024} and \citet{Harvey2024}. A notable tension emerges with the \citet{Labbe2023} results surpassing the maximum achievable stellar mass density in the $\Lambda$CDM framework. However, EDE exhibits no such challenge, with all data points comfortably below the maximum threshold at the 1-$\sigma$ level. It is feasible to produce these massive galaxies with a physically realistic SFE. The stellar mass density produced by our fiducial star-formation model (with $A(z)$ chosen to match the UVLF at the corresponding redshift) in EDE is marginally consistent with findings from \citet{Wang2024-RUBIES} and \citet{Desprez2024} at $z\simeq 8$ (who revisited the CEERS sample and tested removing the single most massive outlier), and the Cosmos-Web results at $z\sim 10$~\citep{Casey2024}. However, discrepancies persist at $z\simeq 12$. We explore two potential model variations, increasing $A(z)$ such that $\sigma_{\rm UV}(10^{10.5})\simeq 1.3$ mag or increasing the SFE by five times. As shown in Figure~\ref{fig:parameters}, these two can make the UVLF at $z\simeq 16$ consistent with observations. Since the impact of UV variability on $\rho_{\ast}$ is primarily through $\mathcal{F}(\sigma_{\rm sfr})$, it is very hard to reconcile observations purely on the UV variability direction. Alternatively, a heightened SFE emerges as a potential solution for reconciling the model with observational discrepancies in this context, as it not only increases the $M_{\ast}$ at a given $M_{\rm halo}$ but also lowers the integration range in Equation~\ref{eq:rhostar}.

Many massive galaxy candidates mentioned above belong to the population of compact and extremely red galaxies found by JWST at $z\gtrsim 4$, known as the Little Red Dots (LRDs; e.g. \citealt{Matthee2024,Kocevski2024,Kokorev2024}). Similar to what we discussed above, if these LRDs are all interpreted as massive galaxies, the implied stellar mass densities exceed the maximum value allowed by the total baryon content in the Unverse~\citep[e.g.][]{Akins2024}. Since many of them show broad Balmer emission lines as a signature of AGN, a more plausible interpretation of these LRDs is heavily obscured AGN. However, in this scenario, their number densities are still much larger than extrapolated from pre-JWST observations in UV and X-ray~\citep[e.g.][]{Aird2015,Kulkarni2019,Shen2020,Niida2020} and some theoretical model predictions after considering their heavy obscuration. As we shown in Figure~\ref{fig:nhalo}, the increase in halo number density at $z\simeq 6-8$ is about $0.3-0.5$ dex in EDE compared to the standard $\Lambda$CDM cosmology and this level of difference could alleviate these puzzles of the general LRD population as well.

\section{Discussion and Conclusions}
\label{sec:conclusions}

We investigate the UVLF of galaxies at $4\lesssim z \lesssim 16$ in the standard $\Lambda$CDM and EDE cosmologies. The specific EDE model we consider has the implication of solving the Hubble tension by reducing the sound horizon at the CMB epoch. As a byproduct of this, the cosmological parameters are changed in EDE such that the abundance of dark matter haloes is systematically enhanced compared to $\Lambda$CDM at high redshift ($z\gtrsim 8$). We describe an empirical galaxy formation model incorporating SFE and UV variability of galaxies that depends on host halo mass. Our benchmark model, with no redshift dependence of SFE or UV variability, can predict UVLF that is consistent with observations at $4 \lesssim z \lesssim 10$. We find that compared to the model with constant UV variability, the halo mass dependence significantly improves the agreement with the observed UVLFs in the faint end.

At higher redshifts ($z \gtrsim 10$), we examined the necessary model variations to reconcile theoretical predictions with observations. We take the approach of adjusting the normalization of $\sigma_{\rm UV}$-$M_{\rm halo}$ relation and alternatively one can also achieve this by adjusting the SFE. Although a moderate level of enhancement in UV variability is still suggested in EDE, the $\sigma_{\rm UV}$ values are consistent with the predictions from hydrodynamical simulations. On the contrary, in $\Lambda$CDM, substantially higher $\sigma_{\rm UV}$ values are suggested and, in particular at $z\simeq 16$, unphysically high $\sigma_{\rm UV}$ is required. This suggests that EDE may offer a more feasible framework for understanding galaxy formation and evolution during cosmic dawn. 

Furthermore, the enhanced halo abundance in EDE completely resolves the tension between a handful of potentially ultra-massive galaxies and the maximum stellar mass density allowed by cosmology. In the EDE cosmology, the benchmark galaxy formation model is quantitatively consistent with some stringent observational constraints without fine-tuning. 

Due to different levels of $\sigma_{\rm UV}$ inferred, galaxies in the EDE cosmology are hosted by more massive haloes. This difference in host halo mass is reflected in the effective bias of galaxies, which varies in order of magnitudes between models with enhanced SFE, constant $\sigma_{\rm UV}$, halo mass-dependent $\sigma_{\rm UV}$ in different cosmologies. Future observations may be able to distinguish these somewhat degenerated scenarios through clustering measurements. As shown in Figure~\ref{fig:sfe} and Figure~\ref{fig:sigma_vs_mhalo}, cosmological simulations of galaxy formation at high redshifts are still far from reaching converged results in terms of SFE and UV variability. The major uncertainties originate from the explicit modelling of star-formation and multiple channels of stellar feedback in resolved ISM~\citep[e.g.][]{Iyer2020,Zhang2024}, potentially entangled with AGN feedback, variable IMF, etc. JWST observations offer new opportunities to constrain this class of galaxy formation models in simulations, such as emission line measurements~\citep[e.g.][]{Endsley2023,Sun2023-OIII,Boyett2024,Helton2024}, size and morphological constraints~\citep[e.g.][]{Baggen2023,Robertson2023,Morishita2024,deGraaff2024b,Shen2024,Fujimoto2024}, as well as chemical enrichment patterns~\citep[e.g.][]{Bunker2023,Curti2023,Cameron2023}. Narrowing down the uncertainties in SFE and variability and their redshift evolution can help disentangle the solutions through baryonic physics versus cosmology. Meanwhile, direct constraints on EDE and all ``early'' solutions to the Hubble tension can be obtained by independent measurements of the age of the Universe from e.g. globular clusters, white dwarfs, and metal-poor stars in the local Universe~\citep[e.g.][]{Winget1987, Cowan1991, Chaboyer1995, Vandenberg1996, Verde2019}, but systematic uncertainties remain large~\citep[e.g.][]{BK2021,Ying2023}.

Our findings highlight the potential of using observed galaxy abundance at cosmic dawn to constrain cosmological models. Specifically, the EDE model shows promise in providing a unified solution to the Hubble tension and the puzzles of massive, UV-bright galaxies at cosmic dawn. Excitingly, forthcoming CMB experiments focused on TT and EE power spectra on small angular scales have the potential to strongly detect or exclude the presence of EDE, which will offer substantial clarity for both models of cosmology and high-redshift galaxy formation.

\section*{Acknowledgements}
We thank Robert Feldmann, and Stephanie O'Neil for suggestions that improved the manuscript. MV acknowledges support through NASA ATP 19-ATP19-0019, 19-ATP19-0020, 19-ATP19-0167, and NSF grants AST-1814053, AST-1814259, AST-1909831, AST-2007355 and AST-2107724. MBK acknowledges support from NSF CAREER award AST-1752913, NSF grants AST-1910346 and AST-2108962, NASA grant 80NSSC22K0827, and HST-GO-16686, HST-AR-17028, and HST-AR-17043 from the Space Telescope Science Institute (STScI), which is operated by AURA, Inc., under NASA contract NAS5-26555. RPN acknowledges support for this work provided by NASA through the NASA Hubble Fellowship grant HST-HF2-51515.001-A awarded by STScI.

\section*{Data Availability}
The data underlying this article can be shared on reasonable request to the corresponding author. The source code and the observational data compiled for this project are publically available at \href{https://github.com/XuejianShen/highz-empirical-variability}{the online repository}.



\bibliographystyle{mnras}
\bibliography{main} 





\bsp	
\label{lastpage}
\end{document}